\documentstyle[12pt]{article}
 \def\y{y }
\else 

\fi
\renewcommand{\theequation}{\thesection.\arabic{equation}}
\renewcommand{\thefootnote}{\fnsymbol{footnote}}
\newlength{\extraspace}
\newlength{\extraspaces}
\newcommand{\be}{\begin{equation}
\addtolength{\abovedisplayskip}{\extraspaces}
\addtolength{\belowdisplayskip}{\extraspaces}
\addtolength{\abovedisplayshortskip}{\extraspace}
\addtolength{\belowdisplayshortskip}{\extraspace}}
\newcommand{\ee}{\end{equation}}

\newcommand{\ba}{\begin{eqnarray}
\addtolength{\abovedisplayskip}{\extraspaces}
\addtolength{\belowdisplayskip}{\extraspaces}
\addtolength{\abovedisplayshortskip}{\extraspace}
\addtolength{\belowdisplayshortskip}{\extraspace}}
\newcommand{\ea}{\end{eqnarray}}

\newcommand{\bas}{\begin{eqnarray*}
\addtolength{\abovedisplayskip}{\extraspaces}
\addtolength{\belowdisplayskip}{\extraspaces}
\addtolength{\abovedisplayshortskip}{\extraspace}
\addtolength{\belowdisplayshortskip}{\extraspace}}
\newcommand{\eas}{\end{eqnarray*}}

\newcounter{subequation}[equation]
\makeatletter

\expandafter\let\expandafter
\reset@font\csname reset@font\endcsname

\def\subeqnarray{\arraycolsep1pt
    \def\@eqnnum\stepcounter##1{\stepcounter{subequation}%
        {\reset@font\rm(\theequation\alph{subequation})}}
\jot5mm     \eqnarray}

\makeatother

\newcommand{\newappendix}[1]{
\vspace{15mm}
\pagebreak[3]
\addtocounter{section}{1}
\setcounter{equation}{0}
\setcounter{subsection}{0}
\setcounter{footnote}{0}
\renewcommand{\theequation}{\Alph{section}.\arabic{equation}}
\begin{flushleft}
{\large\bf \Alph{section}. #1}
\end{flushleft}
\nopagebreak
\medskip
\nopagebreak}

\def\eql#1{\\(#1)\vadjust{\penalty10000\vskip-5.3ex}}
\def\eqll#1{\\ #1\vadjust{\penalty10000\vskip-5.3ex}}

\newcommand{\newsection}[1]{
\vspace{15mm}
\pagebreak[3]
\addtocounter{section}{1}
\setcounter{equation}{0}
\setcounter{subsection}{0}
\setcounter{footnote}{0}

\begin{flushleft}
{\large\bf \thesection. #1}
\end{flushleft}
\nopagebreak
\medskip
\nopagebreak}

\newcommand{\newsubsection}[1]{
\vspace{1cm}
\pagebreak[3]

\addtocounter{subsection}{1}
\noindent{ \bf \thesection.\arabic{subsection} #1}
\nopagebreak
\vspace{2mm}
\nopagebreak}

\newcommand{\newappsection}[1]{
\vspace{1cm}
\pagebreak[3]

\addtocounter{subsection}{1}
\noindent{ \bf \Alph{section}.\arabic{subsection} #1}
\nopagebreak
\vspace{2mm}
\nopagebreak}

\newcommand{\NP}[1]{Nucl.\ Phys.\ {\bf #1}}
\newcommand{\PL}[1]{Phys.\ Lett.\ {\bf #1}}
\newcommand{\CMP}[1]{Comm.\ Math.\ Phys.\ {\bf #1}}

\newcommand{\IJMP}[1]{Int.\ J.\ Mod.\ Phys.\ {\bf #1}}

\newcommand{\N}{\mbox{I\hspace{-.4ex}N}}
\newcommand{\C}{\mbox{$\,${\sf I}\hspace{-1.2ex}{\bf C}}}
\newcommand{\Z}{\mbox{{\sf Z}\hspace{-1ex}{\sf Z}}}
\newcommand{\R}{\mbox{\rm I\hspace{-.4ex}R}}
\newcommand{\1}{\mbox{1\hspace{-.6ex}1}}
\newcommand{\bra}{\langle}
\newcommand{\ket}{\rangle}
\newcommand{\ra}{\rightarrow}

\newcommand{\rra}{\ \longrightarrow \ }

\newcommand{\is}{ &\!=\!& }
\newcommand{\nonum}{\nonumber \\[1.5mm]}
\newcommand{\sspace}{\makebox[1cm]{ }}
\newcommand{\bspace}{\makebox[2cm]{ }}

\newcommand{\th}{{\theta}}
\newcommand{\lb}{\lambda}

\newcommand{\sbar}{{\overline{s}}}

\newcommand{\vbar}{{\overline{v}}}

\newcommand{\Sbar}{{\overline{S}}}

\newcommand{\cO}{{\cal O}}

\newcommand{\cS}{{\cal S}}

\newcommand{\Zb}{{\overline{Z}}}

\newcommand{\Sigmabar}{{\overline{\Sigma}}}
\newcommand{\e}{\epsilon}
\begin{document}
%
\begin{titlepage}
%
\renewcommand{\thefootnote}{\fnsymbol{footnote}}
\begin{flushright}
MPI-PhT/94-74\\
hep-th/9412166
\end{flushright}
\vspace{15mm}

\begin{center}
{\LARGE $\;\;\;\;\;\;$  An Algebraic Approach to Form Factors}
{\makebox[1cm]{ }
          \\[2cm]
{\large Max R. Niedermaier}\\ [3mm]
{\small\sl Max-Planck-%
Institut f\"{u}r Physik} \\
{-\small\sl Werner Heisenberg Institut - } \\
{\small\sl F\"{o}hringer Ring 6}\\
{\small\sl D-80805 Munich, Germany}}
\vspace{3.5cm}

{\bf Abstract}
\end{center}
\begin{quote}
An associative $*$-algebra is introduced (containing a
$TTR$-algebra as a subalgebra) that implements the form
factor axioms, and hence indirectly the Wightman axioms,
in the following sense: Each $T$-invariant linear functional
over the algebra automatically satisfies all the form
factor axioms. It is argued that this answers the question
(posed in the functional Bethe ansatz) how to select the
dynamically correct representations of the $TTR$-algebra.
Applied to the case of integrable QFTs with diagonal
factorized scattering theory a universal formula for the
eigenvalues of the conserved charges emerges.
\end{quote}
\vfill
\renewcommand{\thefootnote}{\arabic{footnote}}
\setcounter{footnote}{0}
\end{titlepage}
\newsection{Introduction}
There are two major approaches to construct and to solve
integrable Quantum Field Theories (QFTs), the form factor
bootstrap and the Quantum Inverse Scattering Method (QISM).
Both methods are usually considered as being independent,
with characteristic strengths and weaknesses. Let us briefly
recapitulate the essentials of both techniques.

The form factor bootstrap \cite{KarWeisz,Smir} takes an
implementation of the Wightman axioms in terms of form factors
as a starting point (for recent developments see
\cite{Smir1,BeLeCl,BH,Sinh,LeClair,LeClSmir,Luk} and below).
Form factors are matrix elements of local operators between a
multiparticle state and the physical vacuum. As a consequence of
the factorized scattering theory there exists a recursive system
of coupled Riemann-Hilbert equations for these form factors,
which entail that the Wightman functions built from them have
all the required properties. Given a factorized scattering theory
the main problem in this approach
is to identify the operator content of the model and to set
up a correspondence to solutions of the form factor equations.
This requires additional dynamical input. A distinguished
infinite set of local operators are the conserved charges
in involution. Their eigenvalues, once known, can thus serve
to specify the additional dynamical input at least partially.

Usually the generating function for the eigenvalues of the
local conserved charges in an integrable model is computed
by means of the QISM. Principally the QISM
achieves the construction of an integrable lattice model
starting from a given classical integrable field theory
(see \cite{FST,KBI} and references therein).
The dynamics is encoded into a representation of the celebrated
$TTR$ algebra and the QISM gives a prescription how
to determine both, the algebra (i.e. the $R$ matrix) and the
representation class from the classical theory. In particular
the representations relevant for QFTs are constructed as
limits of finite dimensional representations (`continuum limit
of a lattice model'). On each of the finite dimensional
representations the trace $t(\th)$ of the monodromy operator
can be diagonalized by Bethe Ansatz techniques. For the
`correct' $R$ matrix and the `correct' representation the
eigenstates of $t(\theta)$ can then be interpreted
as the asymptotic multi-particle states and the eigenvalues
are generating functions for the eigenvalues of the
conserved charges. From the viewpoint of relativistic QFTs
the main shortcomings of the QISM are:
\begin{itemize}
\item  It does not apply to models where the dynamically
correct representation cannot be built from (algebraic)
Bethe Ansatz techniques. Examples are the chiral sigma models and
the real coupling affine Toda theories.
\item It does not guarantee that a bona-fide relativistic
QFT emerges that satisfies the Wightman axioms.
\end{itemize}
The purpose of this paper is to give an algebraic formulation
of the form factor bootstrap that allows one to address the
diagonalization problem of the conserved charges in the context
of form factors. By the design of the form factors one expects
the above technical shortcomings of the QISM to be absent.
The main ingredient in this algebraic formulation is a doublet of
`form factor' algebras $F_{\pm}(S)$ associated with a given two
particle bootstrap $S$-matrix. It has has the following features:
\begin{itemize}
\item[1.] It applies to any $1+1$ dimensional relativistic QFT
with a mass gap and factorized scattering theory.
\item[2.] It contains an algebra of $TTR$-type as a subalgebra,
where $R$ is the physical $S$ matrix.
\item[3.] It implements the form factor axioms (and hence
indirectly the Wightman axioms) in the sense that any
$T$-invariant linear functional $f^{\pm}$ over $F_{\pm}(S)$ will
automatically solve all the form factor axioms, except for the residue
axioms. The sum $f =f^+ +f^-$ then in addition satisfies the
kinematical residue axiom.
\item[4.] Any $T$-invariant linear functional over $F_{\pm}(S)$
solves a system of linear difference equations
(deformed Knizhnik-Zamolodchikov equations (KZE)).
\end{itemize}
Let us comment on these features. Certain fragments of such
an algebra appeared in the context of Yangian and quantum
double constructions studied by F. Smirnov, D. Bernard and
A. LeClair \cite{Smir1,BeLeCl}. In particular Smirnov considered
realizations of Yangians in terms of vertex operators such that
suitable functionals over the algebra satisfied the deformed KZE
and could asymptotically be set in correspondence to
form factors\cite{Smir2,Smir3,Kyoto1}.
Here we try to separate the algebraic and the representation
theoretical aspects. It is remarkable that an algebra
$F_{\pm}(S)$ exists that implements {\em all}\/ the form factor axioms
for {\em any} choice of representation (generically not of
Fock-type) and for {\em any}\/ factorized scattering theory. The
only condition on the functionals over the algebra needed is
\bas
&& f(X\,T^+(\th)_a^b) = \delta_a^b\,f(X)\nonum
&& f(T^-(\th)_a^b\,X) = \omega(a)\delta_a^b\,f(X)\;,\;\; \th \in \C\;,
\eas
which we refer to as $T$-invariance. Here $X$ is any element
of $F_{\pm}(S)$, the generators being $W^+_a(\th),\;T^{\pm}(\th)_a^b$ and
$W^-_a(\th),\;T^{\pm}(\th)_a^b$, respectively. In the second line
$\omega(a)$ is a phase that depends only on $f$ and $a$ but not on $X$.
For elements $X^{\pm}\in F_{\pm}(S)$ of the form
$X^{\pm}=W^{\pm}_{a_n}(\th_n)\ldots W^{\pm}_{a_1}(\th_1)$ write
$f^{\pm}(X^{\pm}) = f^{\pm}_{a_n\ldots a_1}(\th_n,\ldots,\th_1)$. Given
any doublet of $T$-invariant functionals $f^{\pm}$ over $F_{\pm}(S)$,
respectively, the claim in point three above is that
$$
f_{a_n\ldots a_1}(\th_n,\ldots,\th_1) := f^+(X^+)+f^-(X^-)
$$
satisfies all the form factor axioms, except the
bound state residue axiom. We expect that also the latter
can be implemented algebraically, but since the complete set of
bound state poles is strongly model dependent, this is best
deferred to case studies. Feature three thus means that the
solution of the infinite recursive system of form factor
equations can be replaced by the study of the representation
theory of the algebra $F_{\pm}(S)$. The task of investigating the
representation theory of $F_{\pm}(S)$ should be faciliated by feature
four, since one can exploit the existing body of knowledge on
deformed KZE \cite{FR,qKZ1,qKZ2,Kyoto2}. The second
feature, finally, means that one can address the diagonalization
problem of the trace of the monodromy operator also in
the context of form factors. By
the design of the form factor doublet $F_{\pm}(S)$ the above technical
shortcomings of the QISM should be overcome. In particular, there
seems to be a simple answer to the question \cite{Skly1,Skly2}
how to select the dynamically correct representations of the
$TTR$ (here $TTS$) algebra: The dynamically correct
representations are those that can be extended to $T$-invariant
representations of $F_{\pm}(S)$.
In order to test this criterion we applied it to compute the spectrum
of the conserved charges in QFTs with diagonal factorized scattering
theory. In the case of affine Toda theories we recover our previous
result \cite{MN1}. At least for QFTs with diagonal factorized
scattering theory one thus has available a diagonalization method
independent of, and alternative to, Bethe Ansatz techniques.

The paper is organized as follows. In section 2 we introduce
the form factor doublet $F_{\pm}(S)$ and study some additional
structures on it. The above features of the form factor
algebra are derived in section 3. The criterion how to select
the dynamically correct representation of the $TTR$-algebra and its
application to QFTs with diagonal factorized scattering theory
is discussed in section 4. The algebra $F_{\pm}(S)$ is defined for complex
rapidities and bears no obvious relation to the Zamolodchikov-Faddeev
(ZF) algebra, defined for real rapidities. In the appendix we study
the relation between both algebras and prove the algebraic consistency
of $F_{\pm}(S)$ and a number of related algebras.
\pagebreak

\newsection{The form factor algebra}
\vspace{-1cm}

\newsubsection{Definition of the algebra}

\noindent Let $S_{ab}^{dc}(\th),\;\th\in\C$ be a physical two
particle $S$-matrix i.e. a solution of the following set
of equations. First, the Yang Baxter equation
\be
S_{ab}^{nm}(\th_{12})S_{nc}^{kp}(\th_{13})S_{mp}^{ji}(\th_{23})
=S_{bc}^{nm}(\th_{23})S_{am}^{pi}(\th_{13})S_{pn}^{kj}(\th_{12})\;,
\ee
where $\th_{12}=\th_1-\th_2$ etc. Second,
unitarity (2.2a,b) and crossing invariance (2.2c)
\begin{subeqnarray}
&&S_{ab}^{mn}(\th)\,S_{nm}^{cd}(-\th)=\delta_a^d\delta_b^c\\
&&S_{an}^{mc}(\th)\,S_{bm}^{nd}(2\pi i-\th)=\delta_a^d\delta_b^c\\
&&S_{ab}^{dc}(\th)=C_{aa'}C^{dd'}\,S_{bd'}^{ca'}(i\pi -\th)\;,
\end{subeqnarray}
where (2.2c) together with (2.2a), (2.2b) implies (2.2b), (2.2a),
respectively. Further, real analyticity and bose symmetry
\ba
&& [S_{ab}^{dc}(\th)]^* =S_{ab}^{dc}(-\th^*)\;,\\
&& S_{ab}^{dc}(\th)=S_{ba}^{cd}(\th)\;.
\ea
Finally, the normalization condition
\be
S_{ab}^{dc}(0)=\epsilon_{ab}\delta_a^d\delta_b^c\;,\sspace
\epsilon_{ab}\in\{\pm 1\}\;,\;\;\epsilon_{aa} =-1\;.
\ee
It is convenient to borrow Penrose's abstract index notation
from general relativity \cite{PR}. That is to say, indices
$a,b,\ldots$ are not supposed to take numerical values but merely
indicate the tensorial character of the quantity carrying it.
Vectors $v^a,\,v^b,\ldots$ for example are elements of (classes of)
abstract modules $V^a,\,V^b,\ldots$ of possibly different
dimensionality. Covectors $v_a,\,v_b,\ldots$ are elements of the
dual modules $V_a,\, V_b,\ldots$ and repeated upper and lower
case indices indicate the duality pairing. Indices can be raised
and lowered by means of the constant `charge conjugation matrix'
$C_{ab}$ and its inverse $C^{ab}$, satisfying $C_{ad}C^{db}=
\delta_a^b$.

To any solution of the Yang Baxter eqn.~and the conditions
(2.2), (2.5) consider the associative algebra generated
by $(T^{\pm})_a^b(\th)=:T^{\pm}(\th)_a^b,\;\th\in\C$, and a
unit $\1$ subject to the relations
\vspace{3mm}
\eql{T1}
\bas \jot5mm
&& S_{ab}^{mn}(\th_{12})\,T^{\pm}(\th_2)_n^d T^{\pm}(\th_1)_m^c
=T^{\pm}(\th_1)_a^mT^{\pm}(\th_2)_b^n\,S^{cd}_{mn}(\th_{12})\;,
\nonum
&& S_{ab}^{mn}(\th_{12})\,T^{\pm}(\th_2)_n^d T^{\mp}(\th_1)_m^c
=T^{\mp}(\th_1)_a^mT^{\pm}(\th_2)_b^n\,S^{cd}_{mn}(\th_{12})\;,
\eas
valid for all values of $\th_{12}$. Further
\vspace{3mm}
\eql{T2}
\bas \jot5mm
&& C_{mn}T^{\pm}(\th +i\pi)_a^mT^{\pm}(\th)_b^n=C_{ab}\1\;,\nonum
&& C^{mn}T^{\pm}(\th)_m^aT^{\pm}(\th+i\pi)_n^b=C^{ab}\1\;.
\eas
$T(S)$ can be given the structure of a Hopf algebra with
antipode, comultiplication and counit given by%
\footnote{In particular $s$ is a linear (not antilinear)
anti-homomorphism. The relations (T1), (T2) for $T(S)$
coincide with that of a quantum double in its multiplicative
presentation\cite{qCurr}. The relations (TW) are characteristic for
intertwining operators between quantum double modules. We refrain
from using this language because the relation (S) and its consequences
do not seem to have a natural interpretation in the context of
quantum doubles.}
\ba
&&s T^{\pm}(\th)_a^b =C_{aa'}C^{bb'}\,
T^{\pm}(\th +i\pi)_{b'}^{a'}\;,\nonum
&&\Delta T^{\pm}(\th)_a^b =
T^{\pm}(\th)_a^m \,T^{\pm}(\th)_m^b\;,\nonum
&&\epsilon\, T^{\pm}(\th)_a^b =1\;.
\ea
Now extend the algebra $T(S)$ by generators $W_a(\th),\;0\leq Im\th
\leq 2\pi$ having the follwing linear exchange relations with
$T^{\pm}(\th)_a^b$\hfill
\vspace{4mm}
\eql{TW}
\bas\jot5mm
&& T^{\pm}(\th_0)_a^b\,W_{a_1}(\th_1)=S^{db_1}_{aa_1}(\th_{01})\,
W_{b_1}(\th_1)\,T^{\pm}(\th_0)_d^b\;.
\eas
We remark that no contractions are allowed in these relations
(c.f. appendix A.3). Further impose
\vspace{4mm}
\eql{WW}
\bas \jot5mm
W_a(\th_1)\,W_b(\th_2) \is S^{dc}_{ab}(\th_{12})\;
W_c(\th_2)\,W_d(\th_1) \;,\sspace Re\,\th_{12}\neq 0\;.
\eas
The $W$-generators so far are defined only in the strip $0\leq Im\th
\leq 2\pi$. The extension to other strips
$2\pi k \leq Im\,\th\leq 2\pi(k+1),\;k\in\Z$ is done by repeated use of
the relation
\vspace{4mm}
\eql{S}
\bas\jot5mm
&& C^{mn}\,W_m(\th)T^+(\th +i\pi)_n^a=
C^{mn}\,T^-(\th +i\pi)_n^a W_m(\th +2\pi i)\;,
\eas
which can be viewed as a deformed contracted version of (TW).
Observe that (S) is equivalent to
\ba
&& W_a(\th +2\pi i) = T^-(\th +2\pi i)_a^m \,W_n(\th)\,
                    sT^+(\th)_m^n\;,\nonum
&& W_a(\th -2\pi i) = sT^-(\th -2\pi i)_m^n \,W_n(\th)\,
                    T^+(\th-2\pi i)_a^m\;.
\ea
On the $W$ generators the analytic continuation $\th \ra \th +2\pi i$
is thus implemented by an inner automorphism of the algebra. (Whereas
only Lorentz boosts with real rapidities are unitarily implemented via
the 1+1 dim. Poincar\'e group.)
In summary, for any solution $S$ of equations (2.1)--(2.5) we define an
associative algebra $F_*(S)$ with generators $W_a(\th),\;
(T^{\pm})_a^b(\th)=:T^{\pm}(\th)_a^b,\;\th\in \C$,
a unit $\1$ and the generators $P_{\mu}\;,\epsilon_{\mu\nu}K$ of the
1+1 dimensional Poincar\'e algebra. Except for $P_{\mu}$
all generators transform as scalars under the action
of the Poincar\'e group. The defining relations  of $F_*(S)$ then
are that of $T(S)$ together with (TW), (WW) and (S).

The product of $W$ generators in $F_*(S)$ is defined only when all
relative
rapidities have a nonvanishing real part. For relative rapidities that
are purely imaginary, the product of $W$-generators contains simple poles.
In particular $W_a(\th_1)W_b(\th_2)$ contains a simple pole at
$\th_{12} =\pm i\pi$. The algebra  $F_*(S)$ in which the $W$-generators
in addition satisfy the relation (R$\pm$) below will be denoted by
$F_{\pm}(S)$, respectively. It is convenient to use different symbols
$W^+_a(\th)$ and $W^-_a(\th)$ for $W$-generators satisfying
(R$+$) and (R$-$), respectively. The residue conditions then read
\vspace{4mm}
\eql{R$\pm$}
\bas \jot5mm
&& 2\pi i\,\mbox{res}[W^+_a(\th -i\pi)\,W^+_b(\th)]=- C_{ab}\;,\\
&& 2\pi i\,\mbox{res}[W^-_a(\th +i\pi)\,W^-_b(\th)]=- C_{ab}\;.
\eas
We shall refer to the algebra $F_{\pm}(S)$ as the {\em form factor
doublet}. Let us remark that multiple products of $W$-generators have
been defined only in cases where at most one relative rapidity is purely
imaginary. The extension of the product to cases where two or more
relative rapidities are purely imaginary is tricky and will not be
needed. Implicit in these definitions, of course, is the presupposition
that the above relations define a consistent algebra. The verification
of this fact is deferred to appendix A. The significance of
$F_{\pm}(S)$ in the context of form factors has been outlined in
the introduction and will be detailed in section 3.


\newsubsection{The $*$-operation}

\noindent In technical terms a $*$-operation is an antilinear
anti-involution of some associative algebra. Here we shall denote
such operations by $\sigma$ since $*$ is already used for complex
conjugation. The algebra $F_*(S)$ turns out to admit an antilinear
anti-involution $\sigma$ given by
\be
\sigma(T^{\pm})^b_a(\th) = T^{\mp}(\th^*)_a^b\;,\sspace
\sigma(W_a)(\th) =W_a(\th^*)\;,\sspace\sigma^2 =id\;.
\ee
The same holds for the form factor algebra with
$\sigma(W^{\pm}_a)(\th) =W^{\pm}_a(\th^*)$. The operations:
`application of $\sigma$' and `taking the residue' commute in (R$\pm$).

Usually in a $*$-algebra of bounded operators any
linear form $f$ over the algebra can be used to define a
sesquilinear form contravariant w.r.t. $\sigma$ (the $*$-operation
of the algebra) by $(Y,X):= f(\sigma(Y)X)$. In the case at
hand, the algebra $F_{\pm}(S)$ does not consist of bounded operators
and the usual device to smear the operators with appropriate
test functions is problematic, too. The reason is that we wish
to keep track of the analyticity properties of expressions like
$f(W_{a_1}(\th_1)\ldots W_{a_n}(\th_n))$ as a function of
$\th_1,\ldots,\th_n$. In general such expressions
will be germs of multivalued analytic functions with branch
cuts and singularities, so that the construction of cycles
(in the sense of integration theory) will be a non-trivial
task. Rather than attempting to construct such integration
cycles, it is technically much simpler to generalize the
notion of a linear form instead. Thus we shall consider
linear mappings
$$
f:F_{\pm}(S)\rra G\;,
$$
where $G$ is the space of germs of multivalued analytic
functions in any number of complex variables. Alternatively
one may think of these mappings as linear forms in the
usual sense with the extra condition that the dependence on
the rapidity variables parametrizing the elements of $F_{\pm}(S)$
is locally analytic. We shall refer to such maps as `analytic linear
forms over $F_{\pm}(S)$'. Given any analytic linear form in that sense
one can use the antilinear anti-involution $\sigma$
to define the associated sesquilinear form contravariant w.r.t.
it via
\be
F(Y,\,X) := f(\sigma(Y)\,X)\;.
\ee
\vspace{-1cm}
\newsubsection{Residue equations}

\noindent The relation (S) allows one to compute the residue of
$W^{\pm}_a(\th_1)\,W^{\pm}_b(\th_2)$ for $\th_{12} =\pm i\pi$, given
that at  $\th_{12} =\mp i\pi$ in (R$\pm$). Using (S) and (TW) one finds
\ba
&& 2\pi i\,\mbox{res}[W^+_a(\th +i\pi)\,W^+_b(\th)]=
L^+_{ab}(\th)\;,\nonum
&& 2\pi i\,\mbox{res}[W^-_a(\th -i\pi)\,W^-_b(\th)]=
L^-_{ab}(\th-i\pi)\;,
\ea
where
\ba
&& L^+_{ab}(\th)=C_{mn}\,T^-(\th+i\pi)_a^mT^+(\th)_b^n\;,\nonum
&& L^-_{ab}(\th)=C_{mn}\,T^+(\th)_a^mT^-(\th+i\pi)_b^n\;.
\ea
For $S$-matrices where $S_{ab}^{dc}(\pm i\pi)$ is regular, these
operators satisfy
\be
L^{\pm}_{ab}(\th) =-S_{ab}^{dc}(\pm i\pi)L^{\mp}_{cd}(\th)\;,
\ee
using (T1) and $S_{na}^{bn}(2\pi i)=-\delta_a^b =S_{na}^{bn}(0)$.
In particular, these equations imply that for purely imaginary relative
rapidities the (WW) relations break down for
$W^+_a(\th)$ and $W^-_a(\th)$.

The relations (R$\pm$) and (2.10) also allow one to make contact to a
residue prescription first proposed by Smirnov\cite{Smir2}.
Suppose that in addition to the previous relations one postulates
\ba
W^+_a(\th_1)\,W^-_b(\th_2) \is S^{dc}_{ab}(\th_{12})\;
W^-_c(\th_2)\,W^+_d(\th_1) \;,\sspace Re\,\th_{12}\neq 0\;,
\ea
and
\be
2\pi i\,\mbox{res}[W^{\pm}_a(\th +i\pi)\,W^{\mp}_b(\th)]=0\;.
\ee
The linear combination
$$
W_a(\th):= W^+_a(\th) +W^-_a(\th)
$$
then will again satisfy the relations (TW), (S) and (WW).
For the residues one finds from (R$\pm$) and (2.10), (2.14)
\ba
&& 2\pi i\,\mbox{res}[W_a(\th +i\pi)\,W_b(\th)]=
L^+_{ab}(\th)-C_{ab}\;,\nonum
&& 2\pi i\,\mbox{res}[W_a(\th)\,W_b(\th+i\pi)]=
L^-_{ab}(\th)-C_{ab}\;,
\ea
so that by (2.12) a version of (WW) exchange relations is restored
even (for the residues) at relative rapidities $\pm i\pi$.
Operator-valued residue
equations of the form (2.15) first appeared in \cite{Smir2}.
If one postulates these relations at a fundamental level, however, they
give little insight
"how the $W$-generators manage to have such a residue".
In particular, it would be difficult to construct realizations with this
property. A compelling feature of the relation (S) is that, together with
the simpler numerical residue equations (R$\pm$) and (R),
they imply (2.15).

\newsubsection{$n$-th roots of the {\boldmath $\th \ra \th + n\pi i$}
automorphism}

\noindent The relation (S) also allows one to define a remarkable linear
(not antilinear) anti-homomorphism $\sbar$ on $F_*(S)$. Define
\begin{subeqnarray}
&&\sbar T^{\pm}(\th)_a^b =C_{aa'}C^{bb'}\,
T^{\mp}(\th +i\pi)_{b'}^{a'}\;,\\
&&\sbar W_a(\th) =  W_m(\th)\,sT^+(\th)_a^m=
              sT^-(\th)_a^m\,W_m(\th+2\pi i)\;.
\end{subeqnarray}
Note that the restriction of $\sbar$ to $T(S)$ is not the antipode map
in (2.6), but differs from it by the interchange of $T^+(\th)$
and $T^-(\th)$.
We claim that $\sbar:F_*(S)\ra F_*(S)$ is a linear anti-homomorphism
that squares to the $\th \ra \th +2\pi i$ automorphism i.e.
\be
\sbar^2 T^{\pm}(\th)_a^b =T^{\pm}(\th+2\pi i)_a^b\;,\sspace
\sbar^2(W_a)(\th) = W_a(\th+2\pi i)\;.
\ee
Moreover, the $\sbar$-transformed $W$ generators commute with the original
ones
\be
\sbar W_a(\th_1) W_b(\th_2) =  W_b(\th_2)\sbar W_a(\th_1)\;,
\ee
and may be viewed as an algebraic analogue of `screening operators'.

The equations (2.17), (2.18) follow directly from the definition.
Observe that (2.18) can also be rewritten in the form
\be
W_a(\th_1+2\pi i)W_b(\th_2) = S_{ab}^{dc}(\th_{12}+2\pi i)\,
W_c(\th_2)W_d(\th_1 +2\pi i)\;,
\ee
where equation (2.7) is inserted for $W_a(\th_1+2\pi i)$. The fact
that $\sbar$ is an anti-homomorphism is well-known for the $T(S)$
subalgebra; for the (TW) relations it amounts to
$$
\sbar T^{\pm}(\th_0)_a^b\,\sbar W_{a_1}(\th_1) =
S^{db_1}_{aa_1}(\th_{10})\,
\sbar W_{b_1}(\th_1)\,\sbar T^{\pm}(\th_0)_d^b\;,
$$
which one can verify for both expressions on the r.h.s. of (2.16b).
For the (WW) relations there are correspondingly four cases to be checked.
One finds consistently
$$
\sbar W_a(\th_1)\sbar W_b(\th_2)= S_{ab}^{dc}(\th_{21})
\sbar W_c(\th_2)\sbar W_d(\th_1)\;.
$$
Finally, $\sbar$ acts on (S) as an anti-homomorphism if
$T^-(\th)_a^m \sbar W_m(\th -2\pi i) = \sbar W_m(\th) T^+(\th)_a^m$
holds; which by (S) indeed is an identity. This shows that $\sbar$ is
an anti-homomorphism of $F_*(S)$.

It may be useful to compare $\sbar$ to the adjoint action on a quantum
double. Rewriting the usual definition in terms the generators
$T^{\pm}(\th)_a^b$ one obtains
\be
Ad(T^{\pm}(\th)_a^b)\,X :=
T^{\pm}(\th)_a^d\,X\,sT^{\pm}(\th-2\pi i)_d^b\;,
\ee
where $X$ is itself an element of the double. By means of (T2) the
operation (2.20) has the characteristic properties of an adjoint action.
In the matrix presentation employed here, the same formula can be used
to define an adjoint action of $T(S)$ on the $W$-generators, which in
the context of a double construction play the role of intertwining
operators. The defining relations for such an intertwiner (see e.g.
\cite{FR,BeLeCl}) can be checked to be equivalent to our (TW) relations.
For the above adoint action (TW) implies
\bas
&& Ad(T^{\pm}(\th_0)_a^b)[W_{a_n}(\th_n)\ldots W_{a_1}(\th_1)]\nonum
&&\sspace = S_{aa_n}^{c_nb_n}(\th_{0n})\,
S_{c_na_{n-1}}^{c_{n-1}b_{n-1}}(\th_{0n-1})\;\ldots\;
S_{c_2a_1}^{c_1b_1}(\th_{01})\;W_{b_n}(\th_n)\ldots W_{b_1}(\th_1)\;,
\eas
valid for generic rapidities. Thus, besides being structurally different,
the relation (S) concerns just those contractions where the
(TW) relations break down (c.f. appendix A.3).

Consider now
\begin{subeqnarray}
&& s T^{\pm}(\th)_a^b =C_{aa'}C^{bb'}\,
T^{\pm}(\th +i\pi)_{b'}^{a'}\;,\\
&& sW_a(\th) =  W_m(\th)\,sT^+(\th)_a^m=
              sT^-(\th)_a^m\,W_m(\th+2\pi i)\;.
\end{subeqnarray}
The restriction of $s$ to $T(S)$ is the usual antipode map, while on
the $W$-generators $s$ and $\sbar$ coincide $sW_a(\th) =\sbar W_a(\th)$.
In particular, $s$ again acts as an anti-homomorphism on $T(S)$ and
the (TW) and (WW) relations. The condition that $s$ acts as an
anti-homomorphism on (S) reads
\vspace{4mm}
\eqll{($\mbox{S}^2$)}
\bas \jot5mm
&& T^+(\th+2\pi i)_a^m\, sW_m(\th) =
sW_m(\th +2\pi i)\, T^-(\th +2\pi i)_a^m\;,
\eas
which however is not an identity in $F_*(S)$.
To cure this problem one may
consider the algebra where $(\mbox{S}^2)$ has been added to the defining
relations. Then $s$ is an anti-homomorphism of this modified algebra and
$s^2W_a(\th)$ is consistently defined on it. Explicitely, one finds the
following equivalent expressions for the square
\ba
s^2(W_a)(\th)
 \is T^+(\th +2\pi i)_a^m \,W_n(\th)\,sT^+(\th)_m^n\nonum
 \is W_n(\th+2\pi i)\,sT^+(\th+2\pi i)_m^nT^-(\th +2\pi i)_a^m\nonum
 \is T^+(\th+2\pi i)_a^m sT^-(\th)_m^n\, W_n(\th +2\pi i)\nonum
 \is sT^-(\th +2\pi i)_m^n\, W_n(\th +4\pi i)\,T^-(\th +2\pi i)_a^m\;.
\ea
In fact $W^-_a(\th):= s^2W_a(\th -2\pi i)$ can be viewed as a
`higher order copy' of $W^+_a(\th):=W_a(\th)$. It again satisfies the
relations (TW) and (WW) and
\be
C^{mn}W^-_m(\th) T^-(\th +i\pi)_n^a =
C^{mn} T^+(\th +i\pi)_n^a W^-_m(\th +2\pi i)\;,
\ee
which differs from (S) by the interchange of $T^+(\th)$ and $T^-(\th)$.
(So that $W_a(\th)$ and $s^2W_a(\th -2\pi i)$ can not quite serve to
model the generators $W^{\pm}_a(\th)$ of $F_{\pm}(S)$.) In addition one
has `mixed' (WW) relations
$$
W^+_a(\th_1)\,W^-_b(\th_2) = S^{dc}_{ab}(\th_{12})\;
W^-_c(\th_2)\,W^+_d(\th_1) \;,\sspace Re\,\th_{12}\neq 0\;.
$$
Clearly the above process can be iterated. Supplementing the defining
relations of $F_*(S)$ by  $(\mbox{S}^k),\;k\leq n$ the powers
$s^kW_a(\th),\;k\leq n$ will be well-defined and yield
higher order copies of $W_a(\th)$ and $sW_a(\th)$.
The odd powers commute with the even ones, and the even powers
satisfy (TW), (WW) and either (S) or the flipped version (2.23).
If no further relations are imposed, this process never leads back to the
original generators i.e. $s^{2n}W_a(\th)\neq W_a(\th +n\pi i)$,
$s^{2n+1}W_a(\th)\neq sW_a(\th +n\pi i)$ for all $n>0$. Truncations can be
achieved by imposing extra relations in the $T(S)$ subalgebra. As an
example, suppose that the following extra condition is imposed
\vspace{4mm}
\eql{T3}
\bas \jot5mm
&& C_{mn}T^+(\th+i\pi)_a^m T^-(\th)_b^n =
 C_{mn}T^-(\th+i\pi)_a^m T^+(\th)_b^n \;,\nonum
&& C^{mn}T^+(\th+i\pi)_m^a T^-(\th)_n^b =
 C^{mn}T^-(\th+i\pi)_m^a T^+(\th)_n^b \;,
\eas
where the first and the second line are related by the application
of $s$. The relation (T3) has a number of implications: First, it
implies that the $W$-generators themselves satisfy the flipped (S)
relations (2.23). This can be seen by comparing two different expressions
for $s^3W_a(\th)$. From (2.22) and (T3) one finds
$$
sT^+(\th+2\pi i)_a^mW_m(\th+4\pi i) = s^3W_a(\th) =
W_m(\th+2\pi i) sT^-(\th +2\pi i)_a^m \;,
$$
which implies (2.23) with $W^-_a(\th)$ replaced by $W_a(\th)$.
In particular this has the consequence that $s$ has an inverse. Set
\be
s^{-1}W_a(\th) =  W_m(\th-2\pi i)\,sT^-(\th-2\pi )_a^m=
              sT^+(\th-2\pi i)_a^m\,W_m(\th)\;,
\ee
which is well-defined and can readily be checked to be inverse to $s$.
Comparing with the previous equation then yields
$s^3W_a(\th) =s^{-1}W_a(\th +4\pi i)$ i.e.
\be
s^4W_a(\th) =W_a(\th +4\pi i)\;,
\ee
which is the truncation announced before.

\newsubsection{Relation of \mbox{\boldmath$F_{\pm}(S)$}
to real rapidity algebras}

\noindent The doublet $F_{\pm}(S)$ bears no obvious relation to the
Zamolodchikov-Faddeev algebra $Z(S)$, which encodes the algebraic
features of a factorized scattering theory. The characteristics
of the ZF-algebra are that there are two sets of generators
$Z_a(\th),\;\Zb^a(\th)$, both defined for {\em real} rapidities
only. Of course also the additional generators $T^{\pm}(\th)_a^b$
are lacking. In appendix A we introduce extended ZF-algebras,
where the generators $Z_a(\th),\;\Zb^a(\th)$ are supplemented
by generators $T^{\pm}(\th)_a^b$ and study the relations
(among the $T$'s and the mixed products $ZT,\,\Zb T$) that can
consistently be imposed in such an extended ZF-algebra. In particular
there exists an algebra $TZ(S)$ of that type, which can be
regarded as the symmetry algebra of the factorized scattering
theory. As a by-product one obtains a proof of the consistency of
the algebra $F_*(S)$ and hence of $F_{\pm}(S)$. Recall from section
2.3 that by means of the relation (S) the numerical residue equations
(R$\pm$) imply operator-valued residue equations (2.10). In order to
mimik this effect in a real rapidity algebra, we introduce an algebra
$R(S)$, which in a sense interpolates between the quotients of the
extended ZF-algebra and the form factor algebra.
The generators of $R(S)$ include pairs of $Z_a(\th),\;\Zb^a(\th)$
generators and are defined for $Im\,\th \in\pi\Z$ only. In a slight
abuse of notation we shall still refer to $R(S)$ as a `real rapidity'
algebra. The delta function term in the ZF-algebra becomes
operator-valued by means of the replacement
$$
4\pi \delta_a^b\,\delta(\th_{12})\1 \rra
[C^{aa'}L^{\pm}_{a' b}(\th_2) -\delta_a^b] \,2\pi \delta(\th_{12})\;.
$$
In appendix A.5 we show that R(S) is a consistently defined associative
extension of the ZF-algebra. Moreover, the generators
$Z_a(\th),\;\Zb^a(\th),\;\th \in i\pi \Z +\R$ can be combined into a
single generator $W_a(\th)$ with complex arguments. The resulting
algebra $F_R(S)$ has generators
$W_a(\th),\;T^{\pm}(\th)_a^b,\;\th \in \C$
and may be viewed as a `reduced' version of a form factor algebra.
The cruical simplification is that the pole singularities in (2.15)
have been replaced by delta function singularities. Off the singularities
$F_R(S)$ is isomorphic to $F_*(S)$. In appendix B we show that $F_R(S)$
can also be considered as arising through a reduction process from an
alternative form factor algebra $F(S)$. Symbolically the relations
among the various algebras are summarized as follows
\bas
&& F(S)\;\stackrel{red.}{\rra}\;F_R(S)\,\simeq\,R(S)\;
 \stackrel{red.}{\rra}\;TZ(S)\;\stackrel{subalg.}{\supset}Z(S)\;.
\eas
The details of this construction are deferred to the appendix.
For the algebraic characterization of form factors described in the
next section only the following relation between the generators
$Z_a(\th),\;\Zb^a(\th),\;\th \in i\pi\Z +\R$ of $R(S)$ and
the generators $W_a(\th)$ of the reduced algebra $F_R(S)$ is needed
\ba
W_a(\th-i\e) \is \left\{\begin{array}{ll}
              Z_a(\th)\;,\bspace                  & \th\in \R\\
              C_{aa'}\Zb^{a'}(\th-i\pi)\;,\sspace & \th\in \R +i\pi\;.
                        \end{array}\right.\nonum
W_a(\th+i\e) \is \left\{\begin{array}{ll}
              C_{aa'}\Zb^{a'}(\th-i\pi)\;,\sspace & \th\in\R\\
              Z_a(\th)\;,\bspace                  & \th\in\R -i\pi\;,
                        \end{array}\right.
\ea
where the limit $\e\ra 0^+$ is to be taken. Off the singularities this
also gives a correspondence between the generators of $F_R(S)$ and
$F_*(S)$.

\pagebreak
\newsection{Form factors as linear forms over
\mbox{\boldmath$F_{\pm}(S)$}}
The Karowski-Weisz-Smirnov axioms for the form factors
\cite{KarWeisz,Smir} are conveniently listed as follows
\begin{itemize}
\item[(1)] Relation between In and Out states.
\item[(2)] Exchange relation.
\item[(3)] KMS-property.
\item[(4)] Kinematical residue axiom.
\item[(5)] Bound state residue axiom.
\item[(6)] Inner product.
\end{itemize}
We shall discuss these axioms consecutively below along with
their algebraic implementation. Form factors will be identified as
sums of T-invariant linear forms over $F_{\pm}(S)$. Recall from
section 2.2 the notion of an analytic linear form over $F_{\pm}(S)$.
We call a linear form {\em T-invariant} if it satisfies in
addition
\begin{subeqnarray}
&& f(X\,T^+(\th)_a^b) = \delta_a^b\,f^{\pm}(X)\\
&& f(T^-(\th)_a^b\,X) = \omega(a)\,\delta_a^b\,f(X)\;,\;\; \th \in \C\;,
\end{subeqnarray}
where $X\in F_{\pm}(S)$ has rapidities separated from $\th$.
In the second line $\omega(a)$ is a phase $|\omega(a)|=1$ that depends
only on the functionals $f^{\pm}$ and $a$ but not on $X$. The extra
condition that of Lorentz invariance $f(X\,K)=0=f(K\,X)=0$, where $K$
is the generator of Lorentz boosts, will always be understood without
further mentioning. For some
$X^{\pm}=W^{\pm}_{a_n}(\th_n)\ldots W^{\pm}_{a_1}(\th_1)\in F_{\pm}(S)$
we shall also write
\be
f^{\pm}(X^{\pm}) = f^{\pm}_{a_n\ldots a_1}(\th_n,\ldots,\th_1)\;,
\ee
which as a function of the rapidity variables may be viewed
as the germ of some analytic function. The main result of
this section is that, due to the structure of $F_{\pm}(S)$, the
(multivalued) analytic functions arising are precisely
the form factors:
\begin{quote}{\em
For every pair of $T$-invariant linear forms $f^{\pm}$ over $F_{\pm}(S)$,
respectively, their sum $f^+ +f^-$ satisfies axioms (2)--(4). The
associated sesquilinear form (2.9) contravariant w.r.t. $\sigma$
satisfies (1).}
\end{quote}
The dependence on the local operator enters through the specific form of
the functionals $f^{\pm}$. Under these conditions Smirnov's formula
\cite[Eqn.(28)]{Smir} for the inner product (6) just has the status of a
definition. Implicit, however, is the statement that the inner product so
defined coincides with the physical inner product on the
space of scattering states; which is why we included it
as part of the axioms. The only axiom not covered
is that for the residues of the bound state poles. We
expect that also this axiom can be implemented on an
algebraic level by supplementing further relations to
$F_{\pm}(S)$. As the complete set of bound state poles is strongly
model-dependent, the discussion of axiom (5) is best deferred
to case studies.

\newsubsection{Verification of the form factor axioms}

\noindent Here we discuss the algebraic implementation of
axioms (1)--(3) consecutively. It suffices to use $T$-invariant
functionals $f$ over the algebra $F_*(S)$. The inclusion of the
kinematical residue axiom is discussed in section 3.3.

(1) Relation between In and Out states: Let
$(\th'_1,\ldots,\th'_m),\;(b_1,\ldots,b_m)$ be the data of
some asymptotic `out' state (rapidities and quantum numbers)
and similarly $(\th_n,\ldots,\th_1)$, $(a_n,\ldots,a_1)$
that of an `in' state. Let
$(F^{\cO})_{a_n\ldots a_1}^{b_1\ldots b_m}
({\th'}^*_1,\ldots,{\th'}^*_m|\th_n,\ldots,\th_1)$
denote the matrix element of some local operator $\cO(x)$
between these states. The axiom (1) states that this matrix
element is related to an $n+m$ particle form factor by
means of the relation%
\footnote{The axiom is usually formulated for real rapidities
only. We consider rapidities in an $i\e$-neighbourhood of the
real axis to emphasize the structure of the underlying involution.}
\ba
&& (F^{\cO})_{a_n\ldots a_1}^{b_1\ldots b_m}
({\th'}^*_1,\ldots,{\th'}^*_m|\th_n,\ldots,\th_1)\nonum
&& = C^{b_1c_1}\ldots C^{b_mc_m}
f^{\cO}_{c_1\ldots c_m a_n\ldots a_1}
(\th'_1-i\pi,\ldots ,\th'_m -i\pi,\th_n,\ldots,\th_1)\;,
\ea
where all rapidities $\th_i,\th_i'$ lie in $\R+i\e$ and are separated.
Here a set of rapidities $(\omega_1,\ldots,\omega_n)$ is called
separated if $|\omega_i -\omega_j|>\delta$, for some $\delta >0$.
Within the algebraic formulation this statement is simply the
definition of the sesquilinear form (2.9) contravariant w.r.t. the
antilinear anti-involution $\sigma$. One has
\ba
&&  f\left(W_{c_1}(\th'_1 -i\pi)\ldots W_{c_m}(\th'_m -i\pi)\;
W_{a_n}(\th_n)\ldots W_{a_1}(\th_1)\right)\nonum
&& = f\left(\sigma(W_{c_1})({\th'}^*_1 +i\pi)\ldots
\sigma(W_{c_m})({\th'}^*_m +i\pi)\;
W_{a_n}(\th_n)\ldots W_{a_1}(\th_1)\right)\nonum
&& = F\left(W_{c_m}({\th'}^*_m +i\pi)\ldots
W_{c_1}({\th'}^*_1 +i\pi)\;,\;
W_{a_n}(\th_n)\ldots W_{a_1}(\th_1)\right)\;.
\ea
Since all rapidities are in an $i\e$-neighbourhood of the real axis
and separated one can also to rewrite (3.4) in terms of the generators
of the algebra $R(S)$ described in appendix A. Write
\bas
&& F_{a_n\ldots a_1}^{b_1\ldots b_m}
(\th^*_1,\ldots,\th^*_m|\th'_n,\ldots,\th'_1)= F(Y,X)\;,
\eas
where $X=Z_{a_n}(\th_n)\ldots Z_{a_1}(\th_1)$,
$Y=\Zb^{b_m}({\th'}^*_m)\ldots \Zb^{b_1}({\th'}^*_1)$. Using the
correspondence (2.26) for $\th'_i\in \R +i\e$ equation (3.4) becomes
\ba
&& F_{a_n\ldots a_1}^{b_1\ldots b_m}({\th'}^*_1,\ldots,{\th'}^*_m|
\th_n,\ldots,\th_1)= F(Y,X)=
f(\sigma(Y)\,X)=\nonum
&& = C^{b_1c_1}\ldots C^{b_mc_m}
f_{c_1\ldots c_m a_n\ldots a_1}
(\th'_1-i\pi,\ldots ,\th'_m -i\pi,\th_n,\ldots,\th_1)\;,
\ea
which is (3.3). The dependence on the local operator $\cO(x)$
enters through the specific form of the functional $f$ and
will in general be suppressed in the notation. In order to
make contact with the usual interpretation of $F(Y,X)$ as a
matrix element, let us assume (if only for heuristic reasons) that
some analogue of the GNS construction exists for the functionals
$f$. (They are not positive functionals and $F_*(S)$ is not a $C^*$
algebra.) Explicitely, suppose that $f$ can be written as a (bi-)
vector functional $f^{\cO}(X) =(|\cO\ket,\,X|v\ket)$, where the
notation $|\cO\ket := \cO(0)|v\ket$ indicates that the
operator $\cO(x)$ is supposed to be uniquely specified by its action
on the vacuum.  The matrix elements and form factors of the
local operator $\cO(x)$ then are related by
\be
F^{\cO}(Y,X) =  (Y|\cO\ket\,,\,X|v\ket) =
(|\cO\ket\,,\,\sigma(Y)\,X|v\ket) = f^{\cO}(\sigma(Y)\,X)\;.
\ee
In particular one sees that for a nontrivial operator $\cO$ one should
take $f(\1)=0$. For the present purposes there there is no need to make
(3.6) precise. The form factors serve only as a tool to construct the
Wightman functions and the usual GNS construction will apply to them.
Here equation (3.6) is merely taken to illustrate that,
compared to the standard definition,  one is dealing with the
form factors of the adjoint operator $\cO^{\dagger}(0)$.
The phase $\omega(a)=\omega(\cO,W_a)$ in (3.1b) represents
the relative locality index of $\cO(x)$ with the asymptotic fields.
The $T$-invariance of $f$ then just corresponds to the
condition that $|v\ket$ and $|\cO\ket$ are `highest weight' states
of the $T(S)$ subalgebra in the sense that
\ba
&& T^+(\th)_a^b|v\ket = \delta_a^b|v\ket\;,\nonum
&& \sigma(T^-)_a^b(\th)|\cO\ket = \omega(a)\,\delta_a^b|\cO\ket\;.
\ea
We shall refer to such representations as $T$-{\em invariant}
representations of $F_*(S)$. Note that it is {\em not} required that
$|v\ket$ and/or $|\cO\ket$ have any vacuum properties w.r.t.~the
$W_a(\th)$ generators; generically one will not be dealing with Fock-type
representations of $F_*(S)$.

(2) Exchange relation: This is a trivial consequence of the
(WW) relations.
\be
f_{a_1\ldots a_{i+1},a_i\ldots a_n}(\th_n\ldots ,\th_i,
\th_{i+1},\ldots \th_1)= S_{a_{i+1}a_i}^{d\;c}(\th_{i+1,i})\;
f_{a_1\ldots c,d\ldots a_n}(\th_n\ldots ,\th_{i+1},
\th_i,\ldots \th_1)\;,
\ee
which is the exchange axiom. Initially this axiom was formulated for
real rapidities only. Here it holds for $Re\,\th_{i+1,i}\neq 0$. This
extension to generic complex rapidities is non-trivial and is equivalent
to a system of linear difference equations (c.f. section 3.2).

(3) KMS property: This axiom prescribes the behaviour of
form factors under analytic continuation $\th \ra \th +2\pi i$
of one of the rapidity variables. The condition is
\be
f_{a_n\ldots a_1}(\th_n +2\pi i,\,\th_{n-1},\ldots ,\th_1)=
\omega(a_n) f_{a_{n-1}\ldots a_1a_n}(\th_{n-1},\ldots ,\th_1,\,\th_n)\;.
\ee
Initially this axiom again was formulated for real rapidities
$\th_1,\ldots,\th_n$ only. We shall see in section 3.2 that it can be
extended to generic complex rapidities.
If one considers $\th$ formally (for the time being) as a
time variable and assumes $\omega(a_n) =1$, this is precisely a KMS
condition for a thermal (quasi) state of inverse temperature $2\pi$.
This analogy can be pushed further\cite{Luk,MN2}, but in the present
context it just serves to motivate the name `KMS property' for (3.9).
By repeated use of the exchange relation (3.8) an equivalent form is
\ba
&&f_{a_n\ldots a_1}(\th_n +2\pi i,\,\th_{n-1},\ldots ,\th_1)
\nonum
&& =\omega(a_n)S_{a_{n-1}c_{n-1}}^{b_{n-1}b_n}(\th_{n-1,n})\,
S_{a_{n-2}c_{n-1}}^{b_{n-2}c_{n-2}}(\th_{n-2,n})\;\ldots\;
S_{a_1a_n}^{b_1c_1}(\th_{1,n})\;
f_{b_n\ldots b_1}(\th_n,\ldots,\th_1)\;.\;
\ea
In the algebraic formulation equation (3.10) is a consequence of
the relations (S). Using the equivalent form (2.7) one has
\ba
&&f\left(W_{a_n}(\th_n+2\pi i)\,W_{a_{n-1}}(\th_{n-1})\ldots
W_{a_1}(\th_1)\right) \nonum
&&\;\; = f\left(T^-(\th_n +2\pi i)_{a_n}^n\,W_m(\th_n)s(T^+)_n^m(\th_n)\,
W_{a_{n-1}}(\th_{n-1})\ldots W_{a_1}(\th_1)\right) \;.
\ea
In the last line one one first applies the vacuum condition (3.1b).
Then one pushes $s(T^{\pm})_{a_n}^m(\th_n)$ to the right,
using (TW) and applies the vacuum condition (3.1a). Comparison yields
(3.10). One can also return to the original form (3.9) i.e.
$$
f\left(W_{a_n}(\th_n +2\pi i)\,W_{a_{n-1}}(\th_{n-1})\ldots
W_{a_1}(\th_1)\right) =\omega(a_n)\, f\left(W_{a_{n-1}}(\th_{n-1})\ldots
W_{a_1}(\th_1)\,W_{a_n}(\th_n)\right)\,.
$$
That is to say, the $T$-invariant functionals $f$, when
restricted to strings of $W_a(\th)$'s automatically are
KMS functionals. The implementation via (3.11) however is more
stringent in that the same $T$-operators also implement the
kinematical residue axiom and satisfy $TTS$ relations.
Moreover, an argument analogous to (3.11) leads to a compatible
system of linear difference equations (deformed KZE)
generalizing (3.10).

\newsubsection{Deformed Knizhnik-Zamolodchikov Equation}

\noindent Identities similar to (3.10) arise when one evaluates
expectation functionals with an insertion of equation (2.7)
at the $i$-th position, using the invariance conditions (3.1a,b).
After some rearrangement one finds
\be
f_{a_n\ldots a_i\ldots a_1}%
(\th_n,\ldots,\th_i+2\pi i,\ldots,\th_1)=
(A_i)_{a_n\ldots a_1}^{b_n\ldots b_1}(\th_n,\ldots,\th_1)\;
f_{b_n\ldots b_1}(\th_n,\ldots,\th_1)\;,
\ee
valid for generic {\em complex} rapidities $\th_n,\ldots,\th_1$. Here
\ba
&\!\!\!\!& (A_i)_{a_n\ldots a_1}^{b_n\ldots b_1}(\th_n,\ldots,\th_1)
\nonum
&\!\!\!\!& \;\;\; =
\omega(a_i)\, S_{a_{i+1}a_i}^{b_{i+1}c_{i+1}}(\th_{i+1,i}-2\pi i)\,
 S_{a_{i+2}c_{i+1}}^{b_{i+1}c_{i+2}}(\th_{i+2,i}-2\pi i)\;
\ldots \;S_{a_nc_{n-1}}^{b_nc_i}(\th_{n,i}-2\pi i)\times\nonum
&\!\!\!\!& \;\;\;\times S_{a_1c_i}^{b_1c_2}(\th_{1,i})\,
S_{a_2c_2}^{b_2c_3}(\th_{2,i})\;\ldots\;
S_{a_{i-1}c_{i-1}}^{b_{i-1}b_i}(\th_{i-1,i})\nonum
&\!\!\!\!& \;\;\; =
\left[\omega(a_i)\,S_{i,i+1}(\th_{i,i+1}+2\pi i)^{-1}\,
S_{i,i+2}(\th_{i,i+2}+2\pi i)^{-1}\;\ldots\;
S_{i,n}(\th_{i,n}+2\pi i)^{-1}\right.\times\nonum
&\!\!\!\!& \;\;\;\times
S_{1,i}(\th_{1,i})\,S_{2,i}(\th_{2,i})\;\ldots\;
S_{i-1,i}(\th_{i-1,i})\Big]_{\;a_n\ldots a_1}^{\;b_n\ldots b_1}\;,
\ea
using the standard matrix notation in the last line. Introducing
the shift operators $(T_if)(\th_n,\ldots,\th_1)=
f(\th_n,\ldots,\th_i +2\pi i,\ldots,\th_1)$  equation (3.12) becomes
a system of linear difference equations
\be
T_if =A_if\;,\sspace 1\leq i\leq n\;,
\ee
of the form studied in \cite{FR} (`deformed Knizhnik-Zamolodchikov
Equations' (KZE)). As such, the operators $A_i$ must satisfy the
compatibility conditions
\be
(T_iA_j)A_i =(T_jA_i)A_j\;.
\ee
This is indeed the case, the computation of \cite[Theorem 5.2]{FR}
carries over. Observe also that the KZE imply
\be
f_{a_n\ldots  a_1}(\th_n+2\pi i,\ldots,\th_1+2\pi i)=
f_{a_n\ldots  a_1}(\th_n,\ldots,\th_1)\;.
\ee
The relevance of the deformed KZE to form factors
was first pointed out by Smirnov \cite{Smir3}, although no derivation
was given. The system of compatible equations (3.14) is much stronger
than the KMS condition (3.9), which corresponds to the
case $i=n$ and {\em real} rapidities $\th_n,\ldots,\th_1$.
Initially both, the exchange relations (3.8) and the KMS condition (3.9),
were introduced for real rapidities only. If one formally applies (3.8)
also for generic complex rapidities (and assumes that $\omega(a_i)$ is
independent of $i$), the deformed KZE can be seen to be equivalent to
(3.10), now valid for generic complex rapidities.
Thus, in upshot, what the deformed KZE tell is that on the
level of the form factors one can extend both, the exchange relations
and the KMS property, to generic complex rapidities. Implicit in this
extension there are strong consistency conditions,
which are made manifest in (3.15). In an algebraic formulation the
main problem is to reconcile (WW)
exchange relations for complex rapidities
$W_a(\th_1)W_b(\th_2)=S_{ab}^{dc}(\th_{12})W_c(\th_2)W_d(\th_1)$
with the implementation of the kinematical residue axiom (Eqn. (3.17)
below). Formally one can achieve this by postulating Smirnov's
operator-valued
residue equation (2.15). If one then extends the quantum double by
antiunitary operators $U_{\pi},\,U^{\dagger}_{\pi}$ that act on the
local operator $\cO$ and considers the In-Out axiom (1) as being given,
one can also arrive at the KZE\cite{BeLeCl}. In the present formulation
the only
extra ingredient is the relation (S), which implies both, the KZE and
kinematical residue equations via (R$\pm$) and (2.10). The latter is
detailed in the next section.

\newsubsection{Kinematical residue axiom}

\noindent This axiom states that
form factors $f_{a_n\ldots a_1}(\th_n,\ldots \th_1)$
have simple poles at relative rapidities $i\pi$ with prescribed
residues. Explicitely
\ba
&& 2\pi i\,\mbox{res}\,f_{a_n\ldots a_{i+1}a_i\ldots a_1}
(\th_n,\ldots ,\th_i+i\pi,\th_i, \ldots ,\th_1)\nonum
&& = f_{b_n\ldots b_{i+2}b_{i-1}\ldots b_1}
(\th_n,\ldots,\th_{i+2},\th_{i-1}\ldots,\th_1)\, C_{a_{i+1} c}\times\nonum
&&\times\Bigg\{\omega(a_{i+1})\,
S^{c_{n-1}b_n}_{c_{i+1}a_n}(\th_{in}+2\pi i)\,
S^{c_{n-2}b_{n-1}}_{c_{n-1}a_{n-1}}(\th_{in-1}+2\pi i)
\ldots S^{c b_{i+2}}_{c_{i+2}a_{i+2}}(\th_{ii+2}+2\pi i)\,\times\nonum
&&\times S^{c_{i-1}b_{i-1}}_{a_ia_{i-1}}(\th_{ii-1})\,
S^{c_{i-2}b_{i-2}}_{c_{i-1}a_{i-1}}(\th_{ii-2})
\ldots S^{c_{i+1}b_1}_{c_2a_1}(\th_{i1})\,-\,
\delta_{a_i}^{c}\delta_{a_n}^{b_n}\ldots\delta_{a_{i+2}}^{b_{i+2}}
\delta_{a_{i-1}}^{b_{i-1}}\ldots \delta_{a_1}^{b_1}\Bigg\}\,.
\ea
We shall make use of the following reformulation of the kinematical
residue axiom. Let $f^{\pm}_{a_n\ldots a_1}(\th_n,\ldots, \th_1)$
be solutions of the form factors axioms (1) -- (3), which instead
of (3.17) satisfy the simpler residue conditions
\ba
&\!\!\!\!& 2\pi i\,\mbox{res}\,f^{\pm}_{a_n\ldots a_{i+1}a_i\ldots  a_1}
(\th_n,\ldots ,\th_i \mp i\pi,\th_i , \ldots ,\th_1)\nonum
&&\;\;\; =-C_{a_{i+1} a_i}\,
f^{\pm}_{b_n\ldots b_{i+1}b_{i-2}\ldots b_1}%
(\th_n,\ldots\th_{i+2},\th_{i-1}\ldots ,\th_1)\;.
\ea
Then
\be
f_{a_n\ldots a_1}(\th_n,\ldots, \th_1):=
f^+_{a_n\ldots a_1}(\th_n,\ldots, \th_1) +
f^-_{a_n\ldots a_1}(\th_n,\ldots, \th_1)
\ee
satisfies (3.17) and hence all of the axioms (1) -- (4). This is a
consequence of the KZE. Rewrite  the $f^-$ term as
\begin{samepage}
\bas
&\!\!\!\!\!\!&f^-_{a_n\ldots a_1}%
(\th_n,\ldots,\th_{i+1},\th_i,\ldots,\th_1)\nonum
&\!\!\!\!\!\!&=\omega(a_{i+1})(A_{i+1})^{b_n\ldots b_1}_{a_{n}\ldots a_1}
(\th_n,\ldots, \th_{i+1}-2\pi i,\th_i,\ldots ,\th_1)
f^-_{a_n\ldots a_1}(\th_n,\ldots,\th_{i+1}-2\pi i,\th_i,\ldots,\th_1)\,.
\eas
\end{samepage}
Since%
\footnote{Note that the contraction $S_{ab}^{dc}(-i\pi)C_{dc}=
C_{aa'}S_{bc}^{ca'}(2\pi i)=-C_{ab}$ is regular, even if
$S_{ab}^{dc}(-i\pi)$ is not.}
\vspace{-5mm}
\ba
&\!\!\!\!&(A_{i+1})^{b_n\ldots b_1}_{a_{n}\ldots a_1}
(\th_n,\ldots, \th_i-\pi i,\th_i,\ldots ,\th_1)C_{b_{i+1}b_i}\nonum
&&\;\;\; =
-S^{c_{n-1}b_n}_{c_{i+1}a_n}(\th_{in}+2\pi i)\,
S^{c_{n-2}b_{n-1}}_{c_{n-1}a_{n-1}}(\th_{in-1}+2\pi i)
\ldots S^{c b_{i+2}}_{c_{i+2}a_{i+2}}(\th_{ii+2}+2\pi i)\,\times\nonum
&&\;\;\;\times S^{c_{i-1}b_{i-1}}_{a_ia_{i-1}}(\th_{ii-1})\,
S^{c_{i-2}b_{i-2}}_{c_{i-1}a_{i-1}}(\th_{ii-2})
\ldots S^{c_{i+1}b_1}_{c_2a_1}(\th_{i1})\;,
\ea
the claim follows. Within the algebraic framework we have seen that
any $T$-invariant functional $f^+$ or $f^-$ over $F_+(S)$ or $F_-(S)$,
respecively, automatically satisfies the form factor axioms (1) -- (3).
Thus, given the residue condition (R$\pm$) for the $W^{\pm}_a(\th)$
generators, the functions (3.2) will satisfy (3.18), so that their
sum satisfies the kinematical residue equation. As an aside we
remark that the latter also follows directly from Smirnov's residue
prescription (2.15)
\ba
&\!\!\!\!\!& 2\pi i\,\mbox{res}\,f_{a_n\ldots a_{i+1}a_i\ldots a_1}
(\th_n,\ldots,\th_i +i\pi,\th_i,\ldots,\th_1)\nonum
&&\;\;\; =f\left(W_{a_n}(\th_n)\ldots W_{a_{i+2}}(\th_{i+2})\,
[L^+_{a_ia_{i+1}}(\th_i)-C_{a_ia_{i+1}}]\,W_{a_{i-1}}(\th_{i-1})
\ldots W_{a_1}(\th_1)\right).\;\;
\ea
Inserting $L^+_{ab}(\th)=C_{mn}\,T^-(\th +i\pi)_a^m T^+(\th)_b^n$
and using the $T$-invariance conditions (3.1), equations (3.17) is
easily reproduced by means of the (TW) exchange relations. For the
reasons explained in section 2.4 we prefer the first derivation.

In summary, we have shown that the algebraic structure of the form
factor algebra $F(S)$ encodes all the form factor axioms
(1)--(4) in the sense stated after equation (3.1). Axiom (6) then
has the status of a definition and the inclusion of bound state
poles is a separate issue.

\newsection{Diagonalization of the conserved charges}
The form factor algebra contains a subalgebra $T(S)$ of
`$TTR$-type'. By an algebra of $TTR$-type we mean any
associative algebra with one or more sets of generators
$T(\th)_a^b$ subject to the relations
$$
R_{ab}^{mn}(\th_{12})\,T(\th_2)_n^d T(\th_1)_m^c =
T(\th_1)_a^mT(\th_2)_b^n\,R^{cd}_{mn}(\th_{12})\;,
$$
where the numerical $R$-matrix $R_{ab}^{dc}(\th_{12})$
characterizes the algebra (similar to the structure constants
of a Lie algebra). Such algebras contain an abelian subalgebra
$[t(\th),\,t(\th')]=0,\;\th\neq \th'$ generated by the traces
$t(\th):=T(\th)_a^a$. One is interested in representations where
this abelian subalgebra is diagonalizable. In the context
of integrable relativistic QFTs the basic proposal is that
for the `correct' $R$-matrix and the `correct' representation the
eigenstates of $t(\th)$ can be interpreted as the asymptotic
multiparticle states and the eigenvalues are generating functions
for the spectrum of the local conserved charges on these states.
The problem to be solved is:
\begin{quote}
{\em What $R$-matrix and what representations are relevant for the
description of a given integrable QFT? $(*)$}
\end{quote}
The remarkable achievement of the QISM is that in many cases it
provides a technique to construct both, the $R$ matrix and the
dynamically correct representation, starting from the classical
theory. As indicated in the introduction there are however classes
of integrable QFTs where this technique does not apply. Since the form
factor algebra contains a subalgebra of $TTR$-type it seems natural to
use the relation to form factors to propose an alternative
solution to the problem $(*)$. In this section we shall formulate
such a criterion and use it to derive a formula for the eigenvalues
of the local conserved charges valid for any QFT with a mass gap and
diagonal factorized scattering theory.
\newsubsection{How to select the dynamically correct
representation of the \mbox{\boldmath $TTR$} algebra}
\vspace{-4mm}

\noindent ``The determination of the representation class of the
canonical commutation relations is a dynamical problem''\cite[p.57]{Haag}
At least in an abstract sense this is the source of much of the
nontrivial features of QFT. Models of the Wightman axioms, it seems,
cannot be defined directly, but only through a limit of `regularized'
auxilary systems, each of which violates one or more of the axioms.
One is forced to adopt such painfully indirect procedures because the
representations of the canonical commutation relations that are compatible
with some non-trivial interaction are not of Fock-type (`Haags theorem'),
but are unaccessible otherwise.

In integrable QFTs specifically, this generic problem has a nonlinear
counterpart: The determination of the representation class of
the $TTR$ algebra is a dynamical problem. The analogy between
both problems has first been pointed out by Sklyanin\cite{Skly2} in
the course of generalizing the algebraic Bethe Ansatz. The motivation
to search for such a generalization stems from the following
consideration. The QISM answers the questions $(*)$ in the following
way: The `correct' $R$-matrix is the one obtained by $q$-deforming
the classical $r$-matrix which appears in the Poisson brackets of the
spatial component of the linear system. Given this $R$-matrix one
can construct an integrable lattice model in finite volume in which
the traces $t_{a,L}(\th)$ ($a$ and $L$ indicating the UV and volume
cutoff, respectively) can be diagonalized by means of the algebraic
Bethe Ansatz.
In general however the cutoffs cannot be removed without running
into singularities, which usually reflects the non-ferromagnetic
nature of the physical vacuum. In the representation theoretical
language used before, this means that the infinite dimensional
representation of the $TTR$ algebra, which one tries to construct as a
limit of the finite dimensional representations in the cutoff theory,
is not (the) one which is compatible with the dynamics. Roughly
speaking, the idea of the functional Bethe Ansatz \cite{Skly2,Skly1}
is to work directly with the cutoff-free continum theory and to find
the dynamically correct infinite dimensional representation by
`educated guess'. The amount of guesswork required depends on the
type of model considered. For models with a ferromagnetic vacuum
it is fairly small and the functional Bethe Ansatz is a genuine
alternative to the QISM in the usual sense. For models where the
physical vacum is not ferromagnetic, the amount of guesswork is
considerable\cite{Skly1}.

Here we propose a criterion how to select the dynamically correct
infinite dimensional representation of the $TTR$-algebra. All
the guesswork is concentrated in finding the excitation spectrum
of the model and the associated two-particle bootstrap $S$-matrix.
As a matter of fact this is usually much simpler than to set up
the machinery of (some version of) the QISM. Moreover, compared to
the sitation in the late 70s, when the QISM was developed\cite{FST},
there are meanwhile several non-perturbative techniques to test proposed
bootstrap $S$-matrices. In particular, the thermodynamic Bethe
Ansatz \cite{TDB} and Monte Carlo simulations \cite{LW} may count as
such. Given a reliable candidate for a bootstrap $S$-matrix one
can built the algebra $F_{\pm}(S)$ with subalgebra $T(S)$ of $TTR$-type,
where $R$ is the physical $S$-matrix. The proposed criterion how
to select the dynamically correct infinite dimensional
representation of the $TTR$ (here: $TTS$)-algebra then simply is
\begin{quote}
{\em Criterion: Select those representations of $T(S)$ that
can be extended to a $T$-invariant representation of $F_{\pm}(S)$.}
\end{quote}
The motivation is as follows. By construction, any bilinear form
defined on a pair of such representations will automatically
solve all the form factor axioms. The choice of a specific
representation satisfying the criterion reflects the choice of
a specific local operator. By the design of the form factor
axioms, the resulting Wightman functions have all the desired
properties; in particular they satisfy locality\cite{Smir}. In
other words, the `dynamical correctness'
is expected to be built in. On the other hand the criterion is
also strong enough to determine the spectrum of the conserved
charges at least when the scattering theory is diagonal.

\newsubsection{Diagonal FST: Spectrum of maximal abelian subalgebra}

\noindent Let a factorized scattering theory be given with
a diagonal $S$-matrix $S_{ab}^{dc}(\th)=S_{ab}(\th)\delta_a^d
\delta_b^c$ satisfying the bootstrap equations (displayed at
the end of section A.1). Consider the form factor algebra
specialized to this case. (T1) becomes
\vspace{4mm}
\eql{T1}\hfill
\bas \jot5mm
&& t_a^{\pm}(\th_1)\, t_b^{\pm}(\th_2)=
t_b^{\pm}(\th_2)\, t_a^{\pm}(\th_1)\;,\sspace
t_a^{\pm}(\th_1)\, t_b^{\mp}(\th_2)=
t_b^{\mp}(\th_2)\, t_a^{\pm}(\th_1)\;,
\eas
where \hfill
\be
T^{\pm}(\th)_a^b =t^{\pm}_a(\th)\,\delta_a^b\;,
\sspace \mbox{(no sum)}\;.
\ee
That is to say, $T(S)$ degenerates into a direct product of abelian
algebra generated by the diagonal elements of $T^{\pm}(\th)_a^b$.
Moreover $T(S)$ essentially becomes independent of $S$. Accordingly
the relations (T2) simplify.
\vspace{4mm}
\eql{T2}
\bas \jot5mm
&& t_{\bar{a}}^{\pm}(\th +i\pi)\,t_a^{\pm}(\th)=\1\;,
\eas
using $C_{ab}=\delta_{a\bar{b}}$. The relations (TW) and (S) become
\vspace{4mm}
\eql{TW}
\bas\jot5mm
&& t_a^{\pm}(\th_0)\,W_b(\th_1) = S_{ab}(\th_{01})\,
W_b(\th_1)\,t_a^{\pm}(\th_0)\;,
\eas
\eql{S}
\bas\jot5mm
&& t_a^{\pm}(\th)W_{\bar{a}}(\th-i\pi) =
W_{\bar{a}}(\th+i\pi)t_a^{\mp}(\th)\;.
\eas
The (WW) relations read
\vspace{4mm}
\eql{WW}
\bas \jot5mm
W_a(\th_1)\,W_b(\th_2) \is S_{ab}(\th_{12})\;
W_b(\th_2)\,W_a(\th_1)\;.
\eas
The residue conditions (R$\pm$) are unchanged; the reversed residue
equations (2.10) become
\ba
&& 2\pi i\,\mbox{res}[W^+_a(\th+i\pi)W^+_b(\th)]=
\delta_{\bar{a}b}\,e_b(\th)\;,\nonum
&& 2\pi i\,\mbox{res}[W^-_a(\th-i\pi)W^-_b(\th)]=
\delta_{\bar{a}b}\,e_b(\th-i\pi)\;,
\ea
where $e_a(\th):= t^-_{\bar{a}}(\th+i\pi)\,t^+_a(\th)$ can be checked
to be a Casimir operator of $F_*(S)$. Nevertheless the representations
of interest are {\em not} the ones on which $e_a(\th)$ assumes a
constant value. The $T$-invariant functionals satisfy
\be
f(X\,t^+_a(\th)) = f(X)\;,\sspace
f(t_a^-(\th)\,X) = \omega(a)f(X)\;,\;\; \th \in \C\;,
\ee
for $X\in F(S)$ with rapidities separated from $\th$. Setting $e_a(\th)$
equal to $\pm 1$ amounts to the reduction $t^+_a(\th)=\pm t^-_a(\th)$,
which is not the case of interest in the context of form factors.
The fact that $e_a(\th)$ commutes with the $W$ generators is
visible on the level of form factors, although in a more indirect way.
The general result described in section three that
$F_{\pm}(S)$ implements
the form factor axioms of course carries over to the diagonal case.
In particular, for the kinematical residue axiom one finds
\ba
&& 2\pi i\,\mbox{res}f_{a_n\ldots a_{k+1}a_k\ldots a_1}
(\th_n\ldots\th_k+i\pi,\th_k,\ldots \th_1)\nonum
&& \;\;\; =\delta_{\bar{a}_{k+1},a_k}\bigg[\prod_{j\neq k,k+1}
S_{a_ka_j}(\th_{kj})-1\bigg]\,f_{a_n\ldots a_{k+2}a_{k-1}\ldots a_1}
(\th_n\ldots\th_{k+2},\th_{k-1},\ldots \th_1)\;.
\ea
The fact that all $j\neq k,k+1$ enter on an equal footing is easily
seen to be related to $e_a(\th)$ being central.

Consider now the eigenstates of the abelian algebra
$T(S)$ on a $T$-invariant representation of $F_*(S)$.
{}From the (TW) relations it follows that these are simply strings of
$W$-generators and the spectrum is multiplicative
\be
t^+_a(\th)\;W_{a_n}(\th_n)\ldots W_{a_1}(\th_1)|v\ket
= \left(\prod_{i=1}^n S_{aa_i}(\th-\th_i)\right)\;
W_{a_n}(\th_n)\ldots W_{a_1}(\th_1)|v\ket\;,
\ee
where all rapidities are supposed to be separated. In accordance
with the general principles outlined in the previous section
one may thus identify the states in (4.5) with the asymptotic
multiparticle states of the theory.
On the other hand let $I^{(n)},\;n\in E$ be the local
conserved charges in involution. The integers $n\in E\subset \N$
are can be assumed to coincide with the Lorentz spin of the charges
they are labeling. By definition these charges have an additive
spectrum and one may define the eigenvalues on a $n$-particle
state by
\be
I^{(n)}\;W_{a_n}(\th_n)\ldots W_{a_1}(\th_1)|v\ket
= \left(\sum_i I^{(n)}(a_i)\,e^{\pm \th_i n}\right)\;
W_{a_n}(\th_n)\ldots W_{a_1}(\th_1)|v\ket\;,
\ee
where the sign option is specified below. Comparing eqn.s
(4.5) and (4.6) one expects the logarithm of $t^+_a(\th)$ to be
a generating functional for the conserved charges. Of course
the meaning of the logarithm first has to be specified. Due to
the natural grading by the particle number this is unproblematic.
On the $n$-particle sector $t_a(\th)$ acts as a finite dimensional
(diagonal) matrix and one can define the logarithm of  $t_a(\th)$
simply through the logarithm of this matrix representation. The
relation between $\ln t_a(\th)$ and the conserved charges then is
\be
\ln\,t^+_a(\th)=\pm i\sum_{n\in E}
\left(\frac{e^{\pm \th}}{c}\right)^n\;I^{(n)}(a)\,I^{(n)}\;,
\ee
where $\pm Re(\th-\th_i)>0$, acting on a multiparticle state with
rapidities $\th_1,\ldots,\th_n$ and $c$ is a normalization
constant. In order to check that both sides of (4.7) have the same
action on multiparticle states, a series expansion of
$\ln S_{ab}(\th)$ is required. It is convenient to split off the
sign factor in (2.5) and write $S_{ab}(\th) =\epsilon_{ab}
e^{i\delta_{ab}(\th)}$. On general grounds the (redefined)
scattering phase $\delta_{ab}(\th)$ admits an expansion
\be
\delta_{ab}(\th)=\pm\sum_{n>0}
\frac{d_{ab}(n)}{n}\;e^{\mp n\th}\;,\sspace
\pm Re\,\th >0,\;0\leq Im\,\th <s_0\;,
\ee
where $s_0$ is the position of the first bound state pole
(i.e. $S_{ab}(is)$ is analytic for $0<s<s_0$).
The $S$-matrix bootstrap equations, equation (A.7) in particular,
put constraints on the expansion coefficients
$d_{ab}(n)$. Besides symmetry $d_{ab}(n)=d_{ba}(n)$ one
finds
\ba
&& d_{ab}(n)=(-)^{n+1} d_{a\bar{b}}(n)\;,\nonum
&& d_{ea}(n)e^{i\eta(a)} + d_{eb}(n)e^{i\eta(b)} +
d_{ec}(n)e^{i\eta(c)} =0\;.
\ea
The most prominent (possibly all) solutions are those
descending from Lie algebraic data. In that case the particles
$a=1,\ldots,r$ are associated with the Dynkin diagram of a
simple Lie algebra $g$ and the possible fusing angles are
selected by the condition
\be
\sum_{l=a,b,c}e^{\pm is\eta(l)}\,q_l^{(s)}=0\;,
\ee
where $(q^{(s)}_1,\ldots,q^{(s)}_r)^T$ is the (normalized)
eigenvector of the Cartan matrix with eigenvalue
$2(1-\cos\frac{s\pi}{h})$ ($h$: Coxeter number, $s$: exponent).
The coefficients $d_{ab}(n)$ then take the form
\be
d_{ab}(n) =d_n\,q^{(n)}_a q^{(n)}_b\;,
\ee
for real constants $d_n$. The equations (4.9) are satisfied by means
of $q^{(n)}_{\bar{a}}=(-)^{n+1}q^{(n)}_a$ and (4.10), respectively.
Notice that the coefficients (4.11) vanish unless $n$ is an
exponent of $g$ modulo $h$, so that the summations over $n\in \N$
can replaced by summmations over $n\in E$, where $E$ is the set of
affine exponents. Both sides of (4.7) then indeed have the same
action on multiparticle states, provided one identifies
\be
I^{(n)}(a) = c^{n/2}\sqrt{\frac{d_n}{n}}\,q_a^{(n)}\;,
\sspace n\in E\;,
\ee
using the symmetry $S_{ab}(\th)=S_{ba}(\th)$. Equation (4.12) provides a
universal formula for the eigenvalues of the conserved charges
in any QFT with a mass gap and diagonal factorised scattering theory.
As mentioned in the introduction, in QFTs of that type the
diagonalization techniques based on the algebraic Bethe Ansatz
do not apply. Here the result is based entirely on the
properties of the form factor algebra. Compared to the functional
Bethe Ansatz no guesswork is required beyond what is needed to
find a reliable candidate for the bootstrap $S$-matrix. In the
case of real coupling affine Toda theories the formula (4.12) reads
explicitely \cite{MN1}
\be
I^{(n)}(a) =
\left(\frac{ m^2 e^T}{2\beta^2}\right)^{n/2} \beta
\left[\frac{h \,\sin\frac{\pi n}{2h}B
                           \sin\frac{\pi n}{2h}(2-B)
                      }{4\pi n\,B\,\sin \frac{\pi n}{h}}
           \right]^{1/2}\; q_a^{(n)}\;,\sspace
n\in E \;.
\ee
Here $h$ is the Coxeter number, $\beta$ is the (bare) coupling constant
and $B=\frac{\beta^2/2\pi}{1+\beta^2/4\pi}$ is a nonperturbative
effective coupling. The tadpole function $T$ is the sum of all
connected vacuum diagrams. It depends on the choice of the
renormalization scheme, but the combination $m^2 e^T$
appearing in (4.13) can be seen to be invariant under the
normal ordering renormalization group. In particular the
parameter $c$ in (4.12) is identified as
\be
c=\frac{ m^2 e^T}{2\beta^2}\;.
\ee
A realization of the form factor algebra (for diagonal
factorized scattering theories) in terms of the conserved charges
and its relation to trace functionals is dicussed in \cite{MN2}.


\newsection{Conclusions}
The motivation for introducing an algebraic approach to form
factors has already been outlined in the introduction. In
particular it yields a novel diagonalization technique for the
conserved charges, independent of, and alternative to, the QISM.
This has been elaborated for the case of QFTs with
diagonal factorized scattering theory and we intend to treat
the non-diagonal case elsewhere. From the viewpoint of form
factors it remains to be seen to what extend the algebra $F_{\pm}(S)$
faciliates the explicit construction of form factors. Since a
considerable body of knowledge has been accumulated for the
deformed KZE, their algebraic implementation within
$F_{\pm}(S)$ should allow one to investigate the subclass of solutions
corresponding to form factors. From a conceptual viewpoint, finally,
it would be interesting to see whether the analogy between the
selection of the dynamically correct representations of the
$TTR$-, and that of the canonical commutation relations can be turned into
a correspondence. If so, the unfavourable conclusion usually
drawn from Haag's theorem could be circumvented in the case
of integrable QFTs.
\vspace{1cm}
\pagebreak

\setcounter{section}{0}
\noindent{\large\bf Appendix}
\vspace{-16mm}

\newappendix{The extended ZF-algebra and its ideals}

\noindent In this appendix we study the relation of the form factor
algebra $F_{\pm}(S)$ to the Za\-mo\-lod\-chi\-kov-Faddeev (ZF)-algebra.
The ZF-algebra has two sets of generators $Z_a(\th)$ and $\Zb^a(\th)$,
both defined for real rapidities only. We shall supplement these
generators by operators $T^{\pm}(\th)_a^b$ having linear exchange
relations with the ZF-operators. The resulting `extended ZF-algebra'
contains non-trivial ideals. The structure of these ideals determines the
relations (among the $T$'s and the mixed products $ZT,\Zb T$)
that can consistently be imposed on the enlarged set of generators,
by switching to the appropriate quotient algebra. In section A.4
examples of such consistent quotient algebras are discussed. As
a by-product one obtains a consistency proof for the algebra
$F_*(S)$, and hence of $F_{\pm}(S)$. In section A.5 we study an algebra
$R(S)$ that can be viewed as a simplified version of a form factor
algebra, in which the pole singularities in (2.15) are replaced by
delta function singularities. Off the singularities the generators
of $R(S)$ can be set into correspondence to that of $F_*(S)$. This
correspondence (2.26) is used in the disussion of the form factor axiom
(1) in section 3.

\newappsection{The Zamolodchikov-Faddeev algebra}

\noindent The  Zamolodchikov-Faddeev algebra \cite{ZZ,FST} is
an associative algebra with generators
$Z_a(\th)$, $\Zb^a(\th),\;\th\in\R$ a unit $\1$ and the
generators $P_{\mu},\;\epsilon_{\mu\nu}K$ of the 1+1
dimensional Poincar\'e algebra. The operators $Z_a(\th)$, $\Zb^a(\th)$
transform as scalars under the action of the Poincare'
group. The defining relations are
\begin{subeqnarray}
 Z_a(\th_1)\,Z_b(\th_2) \is S^{dc}_{ab}(\th_{12})\;
                Z_c(\th_2)\,Z_d(\th_1)\\
 \Zb^a(\th_1)\, \Zb^b(\th_2) \is S^{ab}_{dc}(\th_{12})\;
 \Zb^c(\th_2)\, \Zb^d(\th_1) \\
 \Zb^a(\th_1)\, Z_b(\th_2) \is S^{ac}_{db}(\th_{21})\;
 Z_c(\th_2)\, \Zb^d(\th_1) + 4\pi \delta(\th_{12})\delta_b^a\,\1\\
 Z_a(\th_1)\, \Zb^b(\th_2) \is S^{db}_{ac}(\th_{21}+2\pi i)\;
 \Zb^c(\th_2)\, Z_d(\th_1) +4\pi \delta(\th_{12})\delta_a^b\1\;,
\end{subeqnarray}
where $\th_{12}=\th_1 -\th_2$ etc. and $\delta(\th)$ is the
real delta distribution. We again use Penrose's abstract index notation
\cite{PR}. The tensor $S_{ab}^{dc}(\th)$ is subject to a number of
consistency relations. Associativity enforces
the Yang Baxter equation. Consistency upon iteration impose
the unitarity conditions (2.2) and $S_{an}^{nb}(2\pi i) =-\delta_a^b$,
which follows from crossing invariance and (2.5).
In addition we require real analyticity (2.3), while the
condition (2.4) is not needed momentarily. For any solution $S$
of the Yang Baxter equation subject to the relations (2.2), (2.3),
(2.5) we call the associative algebra (A.1) the Zamolodchikov-
Faddeev algebra $Z(S)$ associated with $S$.

One can formally%
\footnote{$Z(S)_C$ has no significance in the context of form factors,
for example because of the delta function-, rather
than pole-singularities.}
also consider the algebra (A.1) with complex
rapidities. Let $Z(S)_C$ denote the `complexified' algebra.
The $Z(S)_C$ algebra admits a one parameter family of antilinear
anti-involutions
\bas
&& \sigma:Z(S)_C\rra Z(S)_C\;,\nonum
&& \sigma^2 =id\;,\sspace \sigma(z)=z^*\;,\;\;z\in\C\;,\nonum
&& \sigma(XY)=\sigma(Y)\sigma(X)\;,
\eas
defined by
\ba
&& \sigma_{\beta}(Z_a)(\th) =C_{aa'}\Zb^{a'}(\th^* -i\beta)\;,
\nonum
&& \sigma_{\beta}(\Zb^{a})(\th) =C^{aa'}Z_{a'}(\th^* -i\beta)\;,
\bspace \beta\in\R
\ea
This follows from (2.3) and the convention $\sigma[S_{ab}^{dc}(\th)]=
[S_{ba}^{cd}(\th)]^*$ which adheres to the abstract index
notation. In addition there is a linear involution $\omega$
given by
\bas
&& \omega(Z_a)(\th) =C_{aa'}\Zb^{a'}(\th -i\pi)\;,
\nonum
&& \omega(\Zb^{a})(\th) =C^{aa'}Z_{a'}(\th +i\pi)\;,
\eas
which is compatible with $\sigma_{\beta}$ in the sense that
$(\sigma_{\beta}\,\omega)(Z(S))=(\omega\sigma_{\beta})(Z(S))$.

Consider now a $Z(S)_C$-module $\Sigma$ which is highest weight
in the following sense. There exists a vector
$|v_{\beta}\ket\in\Sigma$ s.t.
$$
\sigma_{\beta}(Z_a)(\th)|v_{\beta}\ket =0\;\;\;
\Longleftrightarrow \;\;\;
\Zb^a(\th-i\beta)|v_{\beta}\ket =0\;,\sspace \th\in\R\;.
$$
Given the antilinear anti-involution $\sigma_{\beta}$ on $Z(S)$
one can define a sesquilinear form
$(\;,\;)_{\beta}:\Sigma \times \Sigma \ra \C$ contravariant
w.r.t. it, i.e.
\be
\left(Y|v\ket\,,\;X|v\ket\right)_{\beta}=
\left(|v\ket\,,\;\sigma_{\beta}(Y)X|v\ket\right)_{\beta}\;.
\ee
The evaluation is done by means of the exchange relations
$$
\sigma_{\beta}(Z_a)(\th^*_1)\,Z_b(\th_2) =
S_{ba}^{cd}(\th_{12}+i\pi-i\beta)\;Z_c(\th_2)\,
\sigma_{\beta}(Z_d)(\th^*_1)+ 4\pi C_{ab}\delta(\th_{12}-i\beta)\;.
$$
Note however that this sesquilinear form is defined
uniquely (up to a factor) only for real rapidities.
For real rapidities one finds in particular
\ba
&&\left(Z_{b_m}(\omega_m)\ldots Z_{b_1}(\omega_1)|v\ket\,,
\,Z_{a_n}(\th_n)\ldots Z_{a_1}(\th_1)|v\ket\right)_{\beta}=
\delta_{nm}\prod_{i=1}^nC_{a_ib_i}
4\pi \delta(\theta_i-\omega_i-i\beta)\;,\nonum
&&\bspace \mbox{for}\;\;\;\omega_m>\ldots> \omega_1\;,
\;\;\;\th_n>\ldots> \th_1\;.
\ea
The matrix elements for other relative orderings of rapidities
can be found from (A.1a).
Clearly the same matrix elements can be described in terms of the
dual highest weight module $\Sigmabar$ based on a state $|\vbar\ket
=|\vbar_{\beta}\ket$ satisfying
$$
Z_a(\th)|\vbar_{\beta}\ket =0\;\;\Longleftrightarrow \;\;
\sigma_{\beta}(\Zb^a)(\th-i\beta)|\vbar_{\beta}\ket =0\;,
\sspace \th\in\R\;.
$$
The evaluation is then done by means of
$$
Z_a(\th_1)\,\sigma_{\beta}(Z_b)(\th^*_1) =
S_{ab}^{dc}(\th_{12}-i\pi+i\beta)\;\sigma_{\beta}(Z_c)(\th^*_2)
\,Z_d(\th_1)+ 4\pi C_{ab}\delta(\th_{12}+i\beta)\;.
$$
In fact, provided $S$ satisfies the `Bose symmetry' (2.4),
one can consistently identify
\be
\left(Y|v\ket\,,\;X|v\ket\right)_{\beta}=
\left(\sigma(Y^*)|\vbar\ket\,,\;
\sigma(X^*)|\vbar\ket\right)_{2\pi-\beta}\;,
\ee
where on the r.h.s. $\sigma =\sigma_{2\pi -\beta}$ and
$X^*$ equals $X$ except that the rapidities are replaced with
their complex conjugates.

Let $\Sigma$ be the state space of 1+1 dim.~relativistic QFT
and let $\Omega^{\pm}$ be the bijections onto the asymptotic
in/out spaces (M{\o}ller operators)
\bas
\Omega^+:\Sigma \rra \Sigma_{in}\;,\\
\Omega^-:\Sigma \rra \Sigma_{out}\;.
\eas
In general $\Sigma_{in/out}$ are proper subspaces of $\Sigma$
(because of bound states) and assuming weak asymptotic
completeness, both are isomorphic and are related by the
scattering operator
$$
\cS=\Omega^+(\Omega^-)^{-1}:\Sigma_{in}\rra \Sigma_{out}\;.
$$
The spaces $\Sigma_{in/out}$ are Fock spaces graded by the
number operator $\Sigma_{in/out}=\bigoplus_{n\geq 0}
\Sigma^{(n)}_{in/out}$; where $\Sigma_{in/out}^{(n)}$ are
the $n$-particle subspaces. In the absence of particle
production the $S$-operator preserves the grading
$\cS:\Sigma_{in}^{(n)}\ra \Sigma_{out}^{(n)}$. In an
integrable QFT this $n$-particle scattering operator factorizes
into a product of two particle ones. In terms of the
two particle $S$-matrix one can define the Zamolodchikov-%
Faddeev algebra $Z(S)$ and describe the entire scattering
theory in terms of this algebra. $(S,\Sigma)$ is called
a {\em factorised scattering theory} if
\begin{itemize}
\item $\Sigma_{in/out}$ are highest weight modules of $Z(S)$ with
parameter $\beta =0$. Explicitely, there exists a vector
$|v\ket =|v_0\ket \in \Sigma^{(0)}= \Sigma^{(0)}_{in/out}$
s.t. $\Zb^a(\th)|v\ket =0,\;\th\in \R$, where
$\1,\;Z_a(\th),\Zb^a(\th),\;\th\in\R$ are the generators of
the $Z(S)$ algebra. The inner product is given by
\bas
&&\bra v|Z_{b_1}(\omega_1)\ldots Z_{b_m}(\omega_m)\,,
\,Z_{a_n}(\th_n)\ldots Z_{a_1}(\th_1)|v\ket\nonum
&&\sspace :=\left(Z_{b_m}(\omega_m)\ldots Z_{b_1}(\omega_1)|v\ket\,,
\,Z_{a_n}(\th_n)\ldots Z_{a_1}(\th_1)|v\ket\right)_0\;.
\eas
\item $\Sigma^{(n)}_{in}$ and $\Sigma^{(n)}_{out}$
have a basis of momentum eigenstates
\bas
&& Z_{a_n}(\theta_n)\ldots Z_{a_1}(\theta_1)|v\ket\;,\sspace
\theta_{\pi(1)}<\ldots<\theta_{\pi(n)}\;,\\
&& Z_{a_n}(\theta_n)\ldots Z_{a_1}(\theta_1)|v\ket\;,\sspace
\theta_{\pi(1)}>\ldots>\theta_{\pi(n)}\;,
\eas
respectively, for some fixed permutation $\pi\in S_n$.
\end{itemize}
It is easy to see that this definition adheres to the usual picture.
The factorized scattering theory can equivalently be
described in terms of the dual module $\Sigmabar$; the inner products
are related by (A.5). This definition does not refer to
the M\o ller operators. We conjecture however that the
M\o ller operators can be constructed in terms of
the form factor algebra. Notice that the simple prescription
\ba
&& \widetilde{\Omega}^+ Z_{a_n}(\theta_n)\ldots
Z_{a_1}(\theta_1)|v\ket= Z_{a_{\pi(n)}}(\theta_{\pi(n)})\ldots
Z_{a_{\pi(1)}}(\theta_{\pi(1)})|v\ket\;,
\nonum
&& \widetilde{\Omega}^- Z_{a_n}(\theta_n)\ldots
Z_{a_1}(\theta_1)|v\ket= Z_{a_{\pi(1)}}(\theta_{\pi(1)})\ldots
Z_{a_{\pi(n)}}(\theta_{\pi(n)})|v\ket\;,
\ea
for $\theta_{\pi(1)}<\ldots <\theta_{\pi(n)}$ and $\cS =
\widetilde{\Omega}^+(\widetilde{\Omega}^-)^{-1}$ reproduces the
correct $S$-matrix elements.

We call a factorised scattering theory {\em diagonal} if the 2-particle
$S$-matrix is diagonal $S_{ab}^{cd}(\theta)=S_{ab}(\theta)
\delta_a^c\,\delta_b^d$. The charge conjugation matrix becomes
$C_{ab}=\delta_{\bar{a}b}$, where $a\ra \bar{a}$ is an involution
of $\{1,\ldots, \dim V\}$. Hermitian unitarity and the
crossing invariance eqn.~reduce to $S_{ab}(\theta)=S_{ba}(\theta)
=S_{ab}(-\theta)^{-1}=S_{ab}^{*}(-\theta^*)$ and
$S_{ab}(i\pi -\theta) = S_{a\bar{b}}(\theta)$. For a diagonal
scattering theory the additional relation due to the presence of
bound states takes a particularly simple form
\be
S_{da}(\theta +i\eta(a))\,S_{db}(\theta + i\eta(b))\,
S_{dc}(\theta +i\eta(c)) =1 \;,
\ee
where the triplet $(\eta(a),\eta(b),\eta(c))$ is related
to the conventional fusing angles\cite{Fusing}.

\newappsection{The extended ZF-algebra}

\noindent Consider the ZF-algebra (A.1) supplemented by generators
$T^{\pm}(\th)_a^b$ subject to the relations
\ba
&& T^{\pm}(\th_0)_a^b\,Z_{a_1}(\th_1) = S^{db_1}_{aa_1}(\th_{01})\,
Z_{b_1}(\th_1)\,T^{\pm}(\th_0)_d^b\;,\bspace\;\,\nonum
&& T^{\pm}(\th_0)_a^b\,\Zb^{a_1}(\th_1) =
S^{da_1}_{ab_1}(\th_{10}+2\pi i)\,
\Zb^{b_1}(\th_1)\,T^{\pm}(\th_0)_d^b\;,
\ea
(and no others). Denote this algebra by $\widetilde{TZ}(S)$.
There are several motivations for these exchange relations. First,
if one thinks of $T^{\pm}(\th)_a^b$ as the generators of a quantum
double in its multiplicative presentation, the relations (A8) are
characteristic for intertwining operators between quantum double
modules. Equivalently, the relations (A8) are designed such that
the difference of the left- and the right hand sides of the ZF-algebra
generate a tensorial set of ideals in the associative algebra
with generators $T^{\pm}(\th)_a^b$ and $Z_a(\th),\;\Zb^a(\th)$, subject
only to the relations (A.8). A related fact is that the relations (A.8)
are essentially the only ones for which the commutator
$[T^{\pm}(\th)_a^b,\;\cdot\;]$ acts as a derivation on products of
$W$-generators. Finally, combined with (T2), Eqn.~(A.8)
leads to the expression for the adjoint action (2.20).

In the following we will consider the structure of the algebra
$\widetilde{TZ}(S)$ in more detail. The one-parameter family of
antilinear anti-involutions $\sigma_{\beta}$ of the ZF-algebra can
be extended to a two-parameter family on $\widetilde{TZ}(S)$. Ignoring
a trivial overall phase it is given by
\begin{subeqnarray}
&& \sigma_{\alpha,\beta}(T^{\pm})^b_a(\th) =\cos\alpha\;
T^{\pm}(\th^* +i\pi-i\beta)_a^b+i\sin\alpha\;
T^{\mp}(\th^* +i\pi-i\beta)_a^b\;,\\
&& \sigma_{\alpha,\beta}(Z_a)(\th) =C_{aa'}\Zb^{a'}(\th^*-i\beta)\;,
\sspace \alpha,\beta\in\R\;,\sspace \sigma_{\alpha,\beta}^2 =id\;.
\end{subeqnarray}
To verify this first apply $\sigma =\sigma_{\alpha,\beta}$ to
the (reversed) relations (A.8), using only (A.9b). This gives
\bas
&& \sigma(T^{\pm})_a^b(\th_0^*)\,Z_{a_1}(\th_1) =
S^{b_1d}_{a_1a}(\th_{01}+i\pi-i\beta)\,Z_{b_1}(\th_1)\,
\sigma(T^{\pm})_d^b(\th_0^*)\;,\nonum
&& \sigma(T^{\pm})_a^b(\th_0^*)\,\Zb^{a_1}(\th_1) =
S^{a_1d}_{b_1a}(\th_{10}+i\pi +i\beta)\,
\Zb^{b_1}(\th_1)\,\sigma(T^{\pm})_d^b(\th_0^*)\;,
\eas
Consistency with (A.8) requires that $\sigma(T^{\pm})_a^b(\th^*)$
is a linear combination of $T^{\pm}(\th_0+i\pi -i\beta)$ and
$T^{\mp}(\th_0+i\pi -i\beta)$. If $a$ and $b$ denote the
parameters of the linear combination, the condition $\sigma^2 =id$
implies $|a|^2+|b|^2=1$ and $a^*b+a b^*=0$. Ignoring a trivial overall
phase yields (A.9a).

\newappsection{Ideals of the extended ZF-algebra}

\noindent
It turns out that $\widetilde{TZ}(S)$ contains non-trivial
ideals, which can be factored out. Most obvious are the linear
ideals. The generators $T^+(\th)_a^b$ and  $T^-(\th)_a^b$
have by definition identical exchange relations with $Z(S)$ and
hence, unless discriminated otherwise, could be identified by
factoring out the twosided ideal  $T^+(\th)_a^b- c\,T^-(\th)_a^b$.
Consistency with $T(S)$ fixes the constant $c$ to be $\pm 1$.
(In fact, in extending the involution $\sigma$ from $Z(S)$ to
$\widetilde{TZ}(S)$ we already made use of such a procedure.
For any antilinear anti-involution $\sigma T^{\pm}(\th)_a^b$
were observed to satisfy the same exchange relations with
$Z(S)$ as some linear combination of $T^{\pm}(\th +i\pi -i\beta)_a^b$,
so that $\sigma T^{\pm}(\th)_a^b$ could be identified with a
suitable linear combination thereof.)
Before turning to the ideals quadratic in $T^{\pm}(\th)_a^b$,
consider ideals linear in the generators of $Z(S)$ and linear
in $T^{\pm}(\th)_a^b$. Set
\begin{subeqnarray}
&& (S^{\pm\pm})_a(\th)= Z_m(\th)sT^{\pm}(\th)_a^m -
sT^{\pm}(\th)_a^m Z_m(\th +2\pi i)\;,\\
&& (S^{\pm\mp})_a(\th)= Z_m(\th)sT^{\pm}(\th)_a^m -
sT^{\mp}(\th)_a^m Z_m(\th +2\pi i)\;,\\
&&(\Sbar^{\pm\pm})^a(\th) =\Zb^n(\th)\,T^{\pm}(\th +2\pi i)_n^a -
T^{\pm}(\th +2\pi i)_n^a\,\Zb^n(\th +2\pi i)\;,\\
&&(\Sbar^{\pm\mp})^a(\th) =\Zb^n(\th)\,T^{\pm}(\th +2\pi i)_n^a -
T^{\mp}(\th +2\pi i)_n^a\,\Zb^n(\th +2\pi i)\;.
\end{subeqnarray}
{}From (A.1) and (A.8) one finds
\begin{subeqnarray}
&& S_a(\th_1)\, Z_b(\th_2) = Z_b(\th_2)\,S_a(\th_1)\;,\sspace
\Sbar^a(\th_1)\, \Zb^b(\th_2) = \Zb^b(\th_2)\,\Sbar^a(\th_1)\;,\\
&& S_a(\th_1)\, \Zb^b(\th_2) = \Zb^b(\th_2)\,S_a(\th_1)\;,\sspace
\Sbar^a(\th_1)\, Z_b(\th_2) = Z_b(\th_2)\,\Sbar^a(\th_1)\;,
\end{subeqnarray}
where $S_a(\th)$ stands for one of the operators in (A.10a,b) and
$\Sbar^a(\th)$ for one of (A.10c,d). In (A.11b) we
also assumed $\th_1\neq \th_2$, so that extra terms proportional
to $\delta(\th_{12})$ are absent. Similarly one obtains
\ba
&& T^{\pm}(\th_0)^b_a\,S_c(\th_1) = S_{nc}^{bm}(\th_{01})\;
S_m(\th_1)T^{\pm}(\th_0)^n_a \;+\; (TTS)\;,\nonum
&& T^{\pm}(\th_0)^b_a\,\Sbar^c(\th_1) =
S_{mn}^{cb}(\th_{10}+2\pi i)\;\Sbar^m(\th_1)T^{\pm}(\th_0)^n_a
\;+\;(TTS)\;,
\ea
where the symbolic notation $(TTS)$ denotes terms linear in the
generators of $Z(S)$ and the ideals (A.13) given below. From
(A.11), (A.12) one concludes that for generic rapidities all
of the operators $S_a(\th),\;\Sbar^a(\th)$ generate ideals in
the quotient algebra obtained
from $\widetilde{TZ}(S)$ by dividing out the ideals (A.13) but not in
$\widetilde{TZ}(S)$ itself.

\noindent{\em Warning:} Observe that the relations (A.8) can be
rewritten as
\bas
&& sT^{\pm}(\th_0)_a^b\,Z_{a_1}(\th_1)= S_{a_1d}^{b_1 b}(\th_{10})\;
Z_{b_1}(\th_1)\,sT^{\pm}(\th_0)_a^d\;,
\eas
By contraction this seems to imply
\bas
&& Z_m(\th)\,sT^+(\th)_a^m= -sT^-(\th)_a^m\,Z_m(\th)\;,\nonum
&& Z_m(\th+2\pi i)\,sT^+(\th)_a^m= -sT^-(\th)_a^m\,Z_m(\th+2\pi i)\;.
\eas
using $S_{na}^{bn}(2\pi i) =-\delta_a^b=S_{na}^{bn}(0)$. Similarly,
contracting the second relation (A.8) seems to imply
\bas
&&\Zb^n(\th)\,T^{\pm}(\th)_n^a =
-T^{\pm}(\th)_n^a\,\Zb^n(\th)\;,\nonum
&&\Zb^n(\th+2\pi i)\,T^{\pm}(\th)_n^a =
-T^{\mp}(\th)_n^a\,\Zb^n(\th +2\pi i)\;.
\eas
However, neither of these contracted expressions are valid identities
in $\widetilde{TZ}(S)$ or in one of its quotient algebras. The difference
of the left- and the right hand sides do not generate ideals in
$\widetilde{TZ}(S)$ or in one of its quotient algebras. The reason is
that the contracted relations are no longer compatible with the
ZF-algebra.
If one views the ZF-algebra as arising from dividing out an ideal
in the associative algebra generated by $T^{\pm}(\th)_a^b$ and
$Z_a(\th),\;\Zb^a(\th)$, subject only to the relations (A.8),
this incompatibility can be traced back to the following
fact: Because of mixing effects, an invariant subset of a tensorial
set of ideals may, but need not, generate an ideal by itself.
In the case at hand e.g.
$T^{\pm}(\th_0)_e^f[Z_a(\th_1)\,Z_b(\th_2) - S^{dc}_{ab}(\th_{12})\;
Z_c(\th_2)\,Z_d(\th_1)]$ generates a tensorial set of ideals, its
contraction on the $e=a$ index does not. The above result on the
operators $S_a(\th)$ and $\Sbar^a(\th)$ shows that nevertheless
some `deformed' version of the contracted relations (A.8) can
consistently be imposed in a suitable quotient algebra of
$\widetilde{TZ}(S)$.

Consider now the ideals quadratic in  $T^{\pm}(\th)_a^b$.
We claim that each of the tensorial sets
\begin{subeqnarray}
&& S_{ab}^{mn}(\th_{12})\,T^{\pm}(\th_2)_n^d T^{\pm}(\th_1)_m^c-
T^{\pm}(\th_1)_a^mT^{\pm}(\th_2)_b^n\,S^{cd}_{mn}(\th_{12})\\
&& S_{ab}^{mn}(\th_{12})\,T^{\pm}(\th_2)_n^d T^{\mp}(\th_1)_m^c-
T^{\mp}(\th_1)_a^mT^{\pm}(\th_2)_b^n\,S^{cd}_{mn}(\th_{12})
\end{subeqnarray}
generates a twosided ideal in $\widetilde{TZ}(S)$.
Moreover the ideals (A.13) collectively are invariant
under $\sigma$. To verify this let $I[T^{\pm},T^{\pm}],\;
I[T^{\pm},T^{\mp}]$ denote the tensorial sets appearing in
(A.13a,b), respectively. One can then verify by direct
computation that
$I[T^{\pm},T^{\pm}]\,Z(S)=Z(S)\,I[T^{\pm},T^{\pm}]$,
which is claim (A.13a). Since $T^+(\th)_a^b$ and $T^-(\th)_a^b$
have identical exchange relations with $Z(S)$ one can also
substitute $T^{\mp}(\th)_a^b$ for one of the
$T^{\pm}(\th)_a^b$ pairs in $I[T^{\pm},T^{\pm}]$, which
yields (A.13b). For the invariance under $\sigma$ then
note the following fact $(*)$: For any antilinear
anti-involution $\sigma$ of $\widetilde{TZ}(S)$ that
preserves $Z(S)$, if $I(T,T)$ generates an ideal, so does
$\sigma I(T,T)\simeq I(\sigma T^*,\sigma T^*)$, where
$T^*(\th)_a^b := T(\th^*)_a^b$. Here we
use $\simeq$ to indicate equality modulo a relabeling
of rapidities and/or raising and lowering of indices by
means of the charge conjugation matrix. The fact $(*)$
then follows from the property (2.3) of the $S$-matrix.
Applied to the ideal $I[T^{\pm},T^{\pm}]$ and the antilinear
anti-involution (A.9) one concludes that also
\bas
\sigma I[T^{\pm},T^{\pm}] &\simeq&
\cos^2\alpha\,I[T^{\pm},T^{\pm}] + i\cos\alpha\sin\alpha
\left(I[T^{\pm},T^{\mp}]+ I[T^{\mp},T^{\pm}]\right) \nonum
&&-\sin^2\alpha\, I[T^{\mp},T^{\mp}]
\eas
generates an ideal and similarly for $I[T^{\pm},T^{\mp}]$.
As a consistency check note that the traces of the ideals
(A.13) (after contracting with $S^{-1})$ generate the expected
scalar ideals.

There are many more quadratic ideals in $\widetilde{TZ}(S)$.
Set
\ba
&& F^{\pm\pm}(\th)_a^b :=C_{aa'}C^{mn}\,T^{\pm}(\th)_m^{a'}\,
T^{\pm}(\th+i\pi)_n^b\;,\nonum
&& F^{\mp\pm}(\th)_a^b :=C_{aa'}C^{mn}\,T^{\mp}(\th)_m^{a'}\,
T^{\pm}(\th+i\pi)_n^b\;,
\ea
which is a $\sigma$-invariant set of operators.
One checks that even each of the components
$F^{\pm\pm}(\th)_a^b$  and $F^{\mp\pm}(\th)_a^b$
separately generates an ideal i.e.
$$
F(\th)_a^b\,Z(S)=Z(S)\,F(\th)_a^b\;,
$$
where $F(\th)_a^b$ stands for any of the operators (A.14).
In particular the $F(\th)_a^b$'s  collectively
can consistently be taken to define a tensorial set of ideals,
in accordance with the index structure. Notice also that
$F(\th)_a^b-\lb \delta_a^b$ again generate tensorial ideals for
any $\lb\in\C$. For a product of $T$-generators with relative rapidities
equal to $+i\pi$ the situation is more subtle. Define
\ba
&& G^{\pm\pm}(\th_1,\th_2)_a^b := C^{bb'}C_{mn}\,
T^{\pm}(\th_1+i\pi)_a^m\,T^{\pm}(\th_2)_{b'}^n\;,\nonum
&& G^{\mp\pm}(\th_1,\th_2)_a^b := C^{bb'}C_{mn}\,
T^{\mp}(\th_1+i\pi)_a^m\,T^{\pm}(\th_2)_{b'}^n\;,
\ea
which is a $\sigma$-invariant set of operators. Each of them
satisfies the relations
\ba
&&  G(\th_1,\th_2)_a^b\,Z_{a_3}(\th_3)=
S_{a_3d}^{eb}(\th_{32}+\pi i)S_{ae}^{cb_3}(\th_{13}+i\pi)\,
Z_{b_3}(\th_3)\, G(\th_1,\th_2)_c^d\;,\nonum
&&  G(\th_1,\th_2)_a^b\,\Zb^{a_3}(\th_3)=
S_{ed}^{a_3b}(\th_{23}-\pi i)S_{ab_3}^{ce}(\th_{31}+\pi i)\,
\Zb^{b_3}(\th_3)\, G(\th_1,\th_2)_c^d\;.
\ea
If $G(\th_1,\th_2)_a^b$ however were to generate
a tensor ideal, so should $G(\th_1,\th_2)_a^b -\lb\delta_a^b$
for $\lb\neq 0$. The latter requires
\bas
&& S_{a_3d}^{eb}(\th_{32}+i\pi)S_{ae}^{cb_3}(\th_{13}+i\pi)\delta_c^d
=\delta_a^b\delta_{a_3}^{b_3}\;,\\
&& S_{ed}^{a_3b}(\th_{23}-i\pi)S^{ce}_{ab_3}(\th_{31}+i\pi)\delta_c^d
=\delta_a^b\delta^{a_3}_{b_3}\;.
\eas
Clearly this will not be the case for generic arguments. By
fine-tuning the arguments one finds that
\be
G^{\pm\pm}(\th,\th)_a^b -\lb\delta_a^b\1\;,\sspace
G^{\mp\pm}(\th,\th)_a^b -\lb\delta_a^b\1\;,\sspace \lb\in\C\;,
\ee
generate tensorial ideals. Notice however that the traces
$G(\th,\th)_a^a$ do not generate scalar ideals by themselves.%
\footnote{This is not in conflict with the tensorial character
of $G(\th,\th)_a^b$. Because of mixing effects an invariant
subset of a tensorial set of ideals may, but need not generate
an ideal by itself.} Scalar ideals are in fact obtained from the
traces of $G(\th_1,\th_2)_a^b$ for a different fine-tuning of
the arguments in (A.15), but these coincide with the traces of
$F(\th)_a^b$
\be
F^{\pm\pm}(\th)_a^a = G^{\pm\pm}(\th,\th+2\pi i)_a^a\;,\sspace
F^{\mp\pm}(\th)_a^a = G^{\mp\pm}(\th,\th+2\pi i)_a^a\;.
\ee
We expect (but have not proved) that the above provides
a complete list of quadratic ideals in $\widetilde{TZ}(S)$.
We shall not discuss higher order ideals here. It is not hard
to see that bound state poles in the S-matrix will
give rise to cubic relations among the Z(S) generators for
special, fine-tuned triples of rapidities. This will induce
additional consistency relations for the extended ZF-algebra
and will affect the structure of its higher order ideals.
Clearly this will be sensitive to the details of the bound
state structure and has to be discussed for each model
separately.

\newappsection{Some quotient algebras}

\noindent Having identified the ideals of  $\widetilde{TZ}(S)$ one can
obtain consistent quotient algebras by dividing out an appropriate
subset of the ideals. The question what ideals one decides to
divide out, depends on the class of representations of the
quotient algebra one is interested in. In particular, in the
context of form factors one wishes to keep the generators
$T^+(\th)_a^b$ and $T^-(\th)_a^b$ distinct, so that one is not
allowed to divide out the linear ideal $T^+(\th)_a^b \pm T^-(\th)_a^b$
in $\widetilde{TZ}(S)$. Still, one can divide out other ideals and
an immediate corrolary is the consistency of the algebra $F_*(S)$.
To see this, observe that the operators $Z_a(\th)$ generate a
subalgebra of the ZF-algebra. All the results on the structure of
the ideals in the extended ZF-algebra of course carry over to
this subalgebra, extended in a similar fashion. Doing this, the
rapidity variable in $Z_a(\th)$ initially is real, but for the
purely algebraic purposes considered here, it can also be extended to
complex values. One can then divide out the ideals (A.13),
$F^{\pm\pm}(\th)_a^b-\delta_a^b$, $G^{\pm\pm}(\th,\th)_a^b -\delta_a^b$
and $(S^{+-})_a(\th)$. The resulting associative algebra is by
construction consistent and is isomorphic to $F_*(S)$. Clearly, this
consistency is not affected by imposing the residue conditions
(R$\pm$).

As mentioned before, an algebra obtained  by dividing out the linear
ideal $T^+(\th)_a^b \pm T^-(\th)_a^b$ will not be of relevance in the
context of form factors. Nevertheless, one can consider the quotient
algebra obtained by dividing out the maximal (linear and quadratic)
ideal $I_{max}$ in $\widetilde{TZ}(S)$. The resulting algebra
may be viewed as the symmetry algebra of a factorized scattering
and will be denoted by $TZ(S)$, i.e.
\be
TZ(S)= \widetilde{TZ}(S)\Big/\,I_{max}\;.
\ee
By construction, $TZ(S)$ is a consistent associative algebra.
Explicitely, the defining relations are (A.1) for the
$Z(S)$ subalgebra and
\bas
&& S_{12}(\theta_{12})\;T_2(\theta_2)\,T_1(\theta_1) =
T_1(\theta_1)\,T_2(\theta_2)\;S_{12}(\theta_{12})\;,\\
&&C^{mn}T_m^a(\th)\,T_n^b(\th+i\pi) =C^{ab}\1\;,\sspace
C_{mn}T^m_a(\th+i\pi)\,T^n_b(\th) =C_{ab}\1\;,
\eas
for the $T(S)$ subalgebra. The mixed relations are (A.8) with
with $T^{\pm}(\th)_a^b$ replaced by $T_a^b(\th)$ and
\ba
&& C^{mn}\,Z_m(\th)T(\th +i\pi)_n^a =
C^{mn}\,T(\th +i\pi)_n^a Z_m(\th +2\pi i)\;,\nonum
&& \Zb^n(\th)\,T(\th +2\pi i)_n^a =
T(\th +2\pi i)_n^a\,\Zb^n(\th +2\pi i)\;.
\ea
Except for the last relations the algebra $TZ(S)$ also
appeared in \cite{DeVeg}. The algebra $TZ(S)$ still is endowed with a
linear anti-homomorphism $s$ and an antilinear anti-involution
$\sigma_{\beta}$. Both are obtained in the obvious way from
Eqn. (A.10) and (A.9), respectively.

As seen in section A.1, a factorized scattering theory can be
described in terms of Fock-type representations of the $Z(S)$
algebra alone. Having enlarged the $Z(S)$ algebra to $TZ(S)$ it is
natural to extend the previous representations to Fock-type
representations of $TZ(S)$ by requiring the existence of a vector
$|v_0\ket\in \Sigma$ satisfying $\Zb^a(\th)|v_0\ket =0$ and
\be
T_a^b(\th)|v_0\ket =\delta_a^b|v_0\ket\;,\;\;\;
\sigma_0(T_a^b)(\th)|v_0\ket =
T_a^b(\th +i\pi)|v_0\ket= \delta_a^b|v_0\ket\;,\;\;\;
\th\in\R\;.
\ee
The first Eqn.~fixes the action of $T_a^b(\th)$ on
$Z(S)|v_0\ket$. The explicit form is conveniently obtained from the
adjoint
action (2.20). The second Eqn.~(A.21) guarantees the consistency with the
inner product (A.4). From (A.5) one can also work out the description
in terms of the dual modules, which is equivalent and independent.

\newappsection{The real rapidity algebra \mbox{\boldmath $R(S)$}}

\noindent All quotient algebras of $\widetilde{TZ}(S)$ contain the
ZF-algebra as a subalgebra. In this section we study a modification
of such an algebra, where this is no longer the case in that
the coefficient of the delta distribution term becomes operator-valued.
{}From $\widetilde{TZ}(S)$ one divides out the ideals implementing the
relations of $T(S)$ and
$S^{\pm\pm}_a(\th)$, $(\Sbar^{\pm\pm})^a(\th)$. In addition one modifies
the $\delta$-distribution term in the ZF-algebra to become
operator-valued
and proportional to $[C^{aa'}L^{\pm}_{a'b}(\th_2)-\delta_a^b]$.
This can be considered as an implementation of Smirnov's residue
formula (2.15), with the (cruical) simplification that the pole
singularity
has been replaced by a delta function singularity. The resulting algebra
$R(S)$ is a consistent extension of the ZF-algebra and can be viewed as a
reduced version of the alternative form factor algebra $F(S)$ described
in appendix B.

We define the algebra $R(S)$ as an associative algebra with generators
$Z_a(\th),\;\Zb_a(\th)$, $T^{\pm}(\th)_a^b$ for $\th\in i\pi\Z +\R$,
a unit $\1$ and  the generators $P_{\mu},\;\epsilon_{\mu\nu}K$ of the
1+1 dimensional Poincar\'e algebra. In a slight abuse of notation we
shall still call $R(S)$ a `real rapidity algebra'. The defining relations
are that of $T(S)$ (restricted to rapidities in $i\pi \Z +\R$),
together with
\vspace{4mm}
\eqll{${\mbox(TZ)}_R$}
\bas\jot5mm
&& T^{\pm}(\th_0)_a^b\,Z_{a_1}(\th_1) = S^{db_1}_{aa_1}(\th_{01})\,
Z_{b_1}(\th_1)\,T^{\pm}(\th_0)_d^b\;,\\
&& T^{\pm}(\th_0)_a^b\,\Zb^{a_1}(\th_1) =
S^{da_1}_{ab_1}(\th_{10}+2\pi i)\,
\Zb^{b_1}(\th_1)\,T^{\pm}(\th_0)_d^b\;,
\eas
replacing (TW). Further
\vspace{3mm}
\eqll{$\mbox{(S)}_R$}
\bas\jot5mm
&& C^{mn}\,Z_m(\th)T^+(\th +i\pi)_n^a =
C^{mn}\,T^+(\th +i\pi)_n^a Z_m(\th +2\pi i)\;,\nonum
&& \Zb^n(\th)\,T^-(\th +2\pi i)_n^a =
T^-(\th +2\pi i)_n^a\,\Zb^n(\th +2\pi i)\;,
\eas
replacing (S). Finally the (WW) relations get replaced by a
modified ZF-algebra with operator-valued coefficients of the
singular terms
\vspace{4mm}
\eqll{$\mbox{(ZZ)}_R$}
\bas\jot5mm
\;\;Z_a(\th_1)\,Z_b(\th_2) \is S^{dc}_{ab}(\th_{12})\,
                Z_c(\th_2)\,Z_d(\th_1)\\
\;\; \Zb^a(\th_1)\, \Zb^b(\th_2) \is S^{ab}_{dc}(\th_{12})\;
 \Zb^c(\th_2)\, \Zb^d(\th_1) \\
\;\; \Zb^a(\th_1)\, Z_b(\th_2) \is S^{ac}_{db}(\th_{21})\;
 Z_c(\th_2)\, \Zb^d(\th_1)
+[C^{aa'}L^+_{a'b}(\th_2)-\delta_a^b]\,2\pi\delta(\th_{12})\\
\;\; Z_a(\th_1)\, \Zb^b(\th_2) \is S^{db}_{ac}(\th_{21}+2\pi i)\;
 \Zb^c(\th_2)\, Z_d(\th_1)
+[C^{bb'}L^-_{ab'}(\th_1) -\delta_a^b]\,2\pi\delta(\th_{12})\;,
\eas
where all rapidities are real modulo $i\pi$ and $L^{\pm}_{ab}(\th)$
are defined as in (2.11). The delta functions are to be read as
$\delta(\th):=\delta(Re\,\th)\,\delta_{Im\,\th,0}$, where the Kronecker
delta is $2\pi i$-periodic.

The proof that $R(S)$ is a consistently defined associative algebra
does not follow from the previous results. Because of the
operator-valued coefficients of the delta function, $R(S)$
is not built from subalgebras generated by
$T^{\pm}(\th)_a^b$ and $Z_a(\th),\;\Zb^a(\th)$, respectively.
This induces some new features in demonstration of consistency.
It is convenient to split the discussion into the following items.
\begin{itemize}
\item[(a)] Iteration of $\mbox{(ZZ)}_R$ and associativity of
$\mbox{(ZZ)}_R$.
\item[(b)] Consistency of ${\mbox(TZ)}_R$ with $\mbox{(ZZ)}_R$
\item[(c)] Consistency of $\mbox{(S)}_R$ with $\mbox{(ZZ)}_R$
\end{itemize}
(a) Iteration of $\mbox{(ZZ)}_R$: Applying the  $\mbox{(ZZ)}_R$
relations to the product of $Z$-generators on the r.h.s. should reproduce
the l.h.s. For the first two relations in $\mbox{(ZZ)}_R$ this is
trivial, and for the last two it is a consequence of equation (2.12).
Associativity of $\mbox{(ZZ)}_R$: To work out the conditions imposed by
associativity of the multiplication in triple products of ZF-generators
one rearranges both sides of the identity
$(V_1V_2)V_3 =V_1(V_2V_3)$ by repeated use of $\mbox{(ZZ)}_R$ until the
order of the factors is reversed. Here $V_a(\th)$ stands for either
$Z_a(\th)$ or $\Zb^a(\th)$. The terms cubic in $V$ on
both sides are found to coincide by means of the Yang Baxter
equation. For triple products where both $Z_a(\th)$ and $\Zb^a(\th)$
enter there will three additional terms linear in $V$ on
either side. Their matching conditions are found to be equivalent
to
\ba
L^+_{ad}(\th_0)\,Z_{a_1}(\th_1)\,S^{a_1d}_{bc}(\th_{10})\is
S^{db_1}_{ab}(\th_{01}+i\pi)\,Z_{b_1}(\th_1)\,
L^+_{dc}(\th_0)\;,\nonum
L^+_{ad}(\th_0)\,\Zb^{a_1}(\th_1)\,S^{bd}_{a_1c}(\th_{01})\is
S^{db}_{ab_1}(\th_{10}+i\pi)\,\Zb^{b_1}(\th_1)\,
L^+_{dc}(\th_0)\;,\nonum
L^-_{ad}(\th_0)\,Z_{a_1}(\th_1)\,S^{a_1d}_{bc}(\th_{10}-i\pi)\is
S^{db_1}_{ab}(\th_{01})\,Z_{b_1}(\th_1)\,
L^-_{dc}(\th_0)\;,\nonum
L^-_{ad}(\th_0)\,\Zb^{a_1}(\th_1)\,S^{bd}_{a_1c}(\th_{01}-i\pi)\is
S^{db}_{ab_1}(\th_{10})\,\Zb^{b_1}(\th_1)\,
L^-_{dc}(\th_0)\;.
\ea
All of them can be checked to be identities in $R(S)$ and to be
compatible with (2.12).

\noindent (b)
Consistency of ${\mbox(TZ)}_R$ with $\mbox{(ZZ)}_R$: Further
consistency conditions arise if one pushes $T^{\pm}(\th_0)_a^b$
through both sides of the equation ${\mbox(TZ)}_R$.
In technical terms the
relations ${\mbox(ZZ)}_R$ have to correspond to a twosided ideal in the
algebra generated by $T(S)$ and the ${\mbox(TZ)}_R$ relations alone.
Explicitely one finds
\ba
&& S_{ab}^{mn}(\th_{01})\,L^+_{cn}(\th_1)\,
T^{\pm}(\th_0)_m^d = S_{ca}^{mn}(\th_{10}+i\pi)\,
T^{\pm}(\th_0)_n^d\,L^+_{mb}(\th_1)\;,\nonum
&& S_{ab}^{mn}(\th_{01}+i\pi)\,L^-_{cn}(\th_1)\,
T^{\pm}(\th_0)_m^d = S_{ca}^{mn}(\th_{10})\,
T^{\pm}(\th_0)_n^d\,L^-_{mb}(\th_1)\;,
\ea
together with the same equations where $L_{ab}^{\pm}(\th)$ is replaced by
$C_{ab}$. Again all of them can be checked to be identities in $R(S)$.

\noindent (c) Consistency of $\mbox{(S)}_R$ with $\mbox{(ZZ)}_R$.
The claim here is that if one defines
\ba
&& s(Z_a)(\th) := Z_m(\th)\,sT^+(\th)_a^m =
  sT^+(\th)_a^m\,Z_m(\th+2\pi i)\;,\nonum
&& s(\Zb^a)(\th) :=  \Zb^n(\th)T^-(\th+ 2\pi i)_n^a
= T^-(\th+ 2\pi i)_n^a\,\Zb^n(\th+2\pi i)\;,
\ea
both of the expressions on the r.h.s have the same exchange relations
with the ZF-operators. This is indeed the case. For example one finds
consistently
\ba
&& s(Z_a)(\th_1)Z_b(\th_2) = Z_b(\th_2) s(Z_a)(\th_1)\;,\nonum
&& s(Z_a)(\th_1)\Zb^b(\th_2) = \Zb^b(\th_2) s(Z_a)(\th_1)
+[sT^+(\th_1)_a^b-sT^-(\th_1)_a^b]\,2\pi\delta(\th_{12})\;,
\ea
together with two similar eqn.s where the roles of $Z$ and $\Zb$ are
interchanged. This concludes the demonstration of the consistency of
$R(S)$.

The last point (c) also allows one to extend the antipode map
$s$ on $T(S)$ to a linear anti-homomorphism $s$ on $
R(S)$ by taking (A.24)
as its action on the $Z,\,\Zb$-generators. One verifies that $s$ indeed
acts as a linear anti-homomorphism on the ${\mbox(TZ)}_R$, $\mbox{(S)}_R$
and $\mbox{(ZZ)}_R$ relations. From (A.25) one also sees that for
$(\th_{12}\,\mbox{mod}\,2\pi i)\neq 0$ the original $Z,\Zb$ generators
commute with the $s$ transformed ones. Consistent with that one has
$$
s(L^{\pm}_{ab})(\th_1)\,Z_c(\th_2)=
Z_c(\th_2)\,s(L^{\pm}_{ab})(\th_1)\;,\sspace
s(L^{\pm}_{ab})(\th_1)\,\Zb^c(\th_2)=
\Zb^c(\th_2)\,s(L^{\pm}_{ab})(\th_1)\;,
$$
valid for generic rapidities. Finally note that the algebra $TZ(S)$
described in section A.4 can be recovered from $R(S)$ by means of the
reduction
$$
T^+(\th)_a^b =-T^-(\th)_a^b =: T_a^b(\th)\;.
$$
As in the case of the ZF-algebra one can formally
consider $R(S)$ also for complex rapidities. Let  $R(S)_C$
denote the `complexified' algebra $R(S)$. We claim that
$R(S)_C$ is endowed with a one parameter family of antilinear
anti-involution $\sigma_{\beta}$ given by
\ba
&&\sigma_{\beta}(T^{\pm})^b_a(\th) =
T^{\mp}(\th^* +i\pi-i\beta)_a^b\;,\nonum
&& \sigma_{\beta}(Z_a)(\th) =C_{aa'}\Zb^{a'}(\th^*-i\beta)\;,
\sspace \beta\in\R\;,\sspace \sigma_{\beta}^2 =id\;.
\ea
{}From section A.2 it follows that $\sigma_{\beta}$ is an
antilinear anti-involution of the exchange relations $\mbox{(TZ)}_R$.
Further $Z_a(\th)\ra \sigma_{\beta}(Z_a)(\th)$ is an antilinear
anti-involution of the ZF-algebra
(where the coefficients of the singular terms are constant).
Having modified the ZF-algebra as in $\mbox{(ZZ)}_R$, the invariance of
the last two Eqn.s has to be re-examined. The definitions imply
$$
\sigma_{\beta}(L^{\pm}_{ab})(\th) =
L^{\pm}_{ba}(\th^* \mp i\beta)\;,
$$
from which it is easy to check that $\sigma_{\beta}$ acts as an
antilinear anti-involution also on the last two eqn.s $\mbox{(ZZ)}_R$.
It remains to verify the consistency with the relations $\mbox{(S)}_R$.
Indeed one checks that
$$
\sigma_{\beta}(T^+)(\th +i\pi)_n^a\,
\sigma_{\beta}(Z_m)(\th)\,C^{mn}=
\sigma_{\beta}(Z_m)(\th +2\pi i)\,
\sigma_{\beta}(T^+)(\th +i\pi)_n^a\,C^{mn}
$$
holds by means of the second equation $\mbox{(S)}_R$ and vice versa.
Together it follows that (A.26) indeed defines a one parameter
family of antilinear anti-involutions of $R(S)_C$.
Return now to $\mbox{(ZZ)}_R$. In terms of
$\sigma_{\beta}(Z_a)(\th)$ the last two relations can be rewritten as
\ba
\sigma_{\beta}(Z_a)(\th^*_1)\,Z_b(\th_2) \is
S_{ba}^{cd}(\th_{12}+i\pi-i\beta)\;Z_c(\th_2)\,
\sigma_{\beta}(Z_d)(\th^*_1)\nonum
&+&[L^+_{ab}(\th_2)-C_{ab}]\,2\pi\delta(\th_{12}-i\beta)\;,
\nonum
Z_a(\th_1)\,\sigma_{\beta}(Z_b)(\th^*_2) \is
S_{ab}^{dc}(\th_{12}-i\pi+i\beta)\;\sigma_{\beta}(Z_c)(\th^*_2)
\,Z_d(\th_1)\nonum
&+&[L^-_{ab}(\th_1)-C_{ab}]\,2\pi\delta(\th_{12}+i\beta)\;.
\ea
Comparing now (WW) with (A.27) specialized to $\beta =\pi$,
this suggests that one can actually combine both of the generators
$Z_a(\th)$ and $\Zb^a(\th)$ into a single operator $W_a(\th)$. Define
\ba
W_a(\th-i\e) \is \left\{\begin{array}{ll}
              Z_a(\th)\;,\bspace                  & \th\in \R\\
              C_{aa'}\Zb^{a'}(\th-i\pi)\;,\sspace & \th\in \R +i\pi\;.
                        \end{array}\right.\nonum
W_a(\th+i\e) \is \left\{\begin{array}{ll}
              C_{aa'}\Zb^{a'}(\th-i\pi)\;,\sspace & \th\in\R\\
              Z_a(\th)\;,\bspace                  & \th\in\R -i\pi\;,
                        \end{array}\right.
\ea
where the limit $\e\ra 0^+$ is to be taken. One finds that
all of the relations  $\mbox{(ZZ)}_R$ translate into
\ba
W_a(\th_1)\,W_b(\th_2) \is S^{dc}_{ab}(\th_{12})\;
W_c(\th_2)\,W_d(\th_1) +
\left[ L^{\pm}_{ab}(\th_2)- C_{ab}\right] 2\pi\delta(\th_{12}\mp i\pi)\;,
\ea
where $L^{\pm}_{ab}(\th)$ are the same as in (2.11) and the upper/lower
case options correspond to $\pm Im\,\th >0$, respectively. Similarly
both of the relations $\mbox{(TZ)}_R$ translate into  (TW) and
$\mbox{(S)}_R$ translates into
\be
C^{mn}\,W_m(\th)T^{\pm}(\th +i\pi)_n^a =
C^{mn}\,T^{\pm}(\th +i\pi)_n^a W_m(\th +2\pi i)\;,
\ee
where the upper case corresponds to $\th\in \R+i\e,\R+i\e+i\pi$ and
the lower case to $\th \in \R-i\e,\R-i\e -i\pi$.

In summary, by means of the correspondence (A.28), the algebra $R(S)$
is isomorphic to an associtive algebra with generators $W_a(\th),\;
T^{\pm}(\th)_a^b$, subject to the relations of $T(S)$ together with
(TW), (A.29) and (A.30). The $W$-generators initially are defined for
rapidities in an $i\epsilon$-neigh\-bour\-hood of the axis $\R +i\pi \Z$,
but it is easy to see that the algebraic consistency is preserved if the
range of definition is extended to generic complex rapidities. We shall
denote the resulting algebra by $F_R(S)$. The antilinear anti-involution
(A.26) on $R(S)$ induces an antilinear anti-involution on $F_R(S)$ of
the same form as (2.8). For $\th_{12}\neq \pm i\pi$, the extra
term on the r.h.s of (A.29) is absent. One can then also replace (A.30)
by (S), so that for $\th_{12}\neq \pm i\pi$ $F_R(S)$ is isomorphic to
$F_*(S)$. For relative rapidities that are purely imaginary,
the extra term
in (A.29) is reminiscent of Smirnov's residue formula (2.15),
except for the (cruical) difference that the pole singularity has been
replaced by a delta function singularity. It is easy to see that the naive
device to replace $\delta(\th)$ by $1/\th$ or $1/(\th\pm i0)$ would
{\em not} be algebraically consistent. The reason is that for
$\th_{12}\neq \pm i\pi$, a product of $W$-generators has different
exchange relations with $T^{\pm}(\th)_a^b$ as $L^{\pm}_{ab}(\th_2)/
(\th_{12}\pm i\pi)$. The algebraic consistency off the singularity
$\th_{12} =\pm i\pi$ can be restored by introducing an extra generator
$C_{ab}(\th_1,\th_2)$ and leads to an alternative form factor algebra.

\newappendix{Alternative form factor algebra}
{}From a technical viewpoint, the main problem in constructing an
algebra that implements the form factor axioms is to reconcile
the (WW) exchange relations with a residue formula of the type (2.15).
One would like to implement (2.15) in terms of a simpler, numerical
residue condition. One way to achieve this was described in section
2 and lead to the form factor doublet $F_{\pm}(S)$. Here we present
an alternative extension of the algebra $F_*(S)$, which may be viewed
as an extension of the algebra $F_R(S)$ `off the singularity'. The
advantage is that the form factors directly arise from $T$-invariant
functionals over $F(S)$ (instead of a sum of two such functionals as
for $F_{\pm}(S)$). Further, the algebras $F_R(S)$ or $R(S)$ can directly
be recoved by means of a reduction process. An important disadvantage
of $F(S)$ is that the additional generator $C_{ab}(\th_1,\th_2)$ is
difficult to construct in a realization.

\newappsection{Definition of the algebra \mbox{\boldmath $F(S)$}}

\noindent Recall that in $F_*(S)$ the product of $W$ generators
is defined only when all relative rapidities have a nonvanishing real
part. In order to extend the product to cases where one of the relative
rapidities is purely imaginary, a further generator $C_{ab}(\th_1,\th_2)$,
$\th_1,\th_2\in\C$ is needed. The form factor algebra $F(S)$ is defined
to be the associative extension of the algebra $F_*(S)$ by the generator
$C_{ab}(\th_1,\th_2)$, subject to the following relations:
\vspace{4mm}
\eql{E}
\bas \jot5mm
W_a(\th_1)\,W_b(\th_2) = \left\{ \begin{array}{ll}
 E^+_{ab}(\th_1,\th_2)\;,\sspace &
 Re\,\th_{12}\ra 0^+\;,\;\;\;2\pi >Im\th_{12}>0\\
 E^-_{ab}(\th_1,\th_2)\;,\sspace &
 Re\,\th_{12}\ra 0^-\;,\;\;\;-2\pi <Im\th_{12}<0\;,
\end{array}\right.
\eas
where
\ba
&& E^+_{ab}(\th_1,\th_2):= T^-(\th_1)_a^m\,
[T^+(\th_2)_b^n -T^-(\th_2)_b^n]\,C_{nm}(\th_2,\th_1)\;,\nonum
&& E^-_{ab}(\th_1,\th_2):= [T^+(\th_1)_a^m -T^-(\th_1)_a^m]\,
\,T^+(\th_2)_b^n\,C_{nm}(\th_2,\th_1)\;.
\ea
The operator $C_{ab}(\th_1,\th_2)$ describes the singularity
structure in products of $W$ generators. It is defined to have
the following properties:
\vspace{4mm}
\eql{C1}
\bas \jot5mm
&& C_{ab}(\th_1,\th_2)\,W_c(\th_3) =
W_c(\th_3)\,C_{ab}(\th_1,\th_2)\;.
\eas
Further
\vspace{4mm}
\eql{C2}
\bas \jot5mm
&& C_{ab}(\th_1,\th_2) =S_{ab}^{dc}(\th_{21})\,
 C_{cd}(\th_2,\th_1)\;,\sspace \th_{12}\neq \pm i\pi\;,
\eas
where relative rapidities $\pm i\pi$ have to be excluded to
ensure the consistency of (C3) and (C5) below.
\vspace{4mm}
\eql{C3}
\bas \jot5mm
S_{nb}^{dm}(\th_{10})\,T^{\pm}(\th_0)_a^n\,
C_{mc}(\th_1,\th_2) \is  S_{nc}^{dm}(\th_{02}+2\pi i)\,
C_{bm}(\th_1,\th_2)\,T^{\pm}(\th_0)_a^n\;,\nonum
&& \sspace \th_{10}\neq 0,2\pi i,\;\;\th_{20}\neq 0,2\pi i\;,
\eas
At the excluded points the contracted relation (C3) gets modified to
\vspace{4mm}
\eql{C4}
\bas \jot5mm
&& T^-(\th_1)_a^m\,C_{mb}(\th_1,\th_2)
=C_{bm}(\th_2,\th_1+2\pi i)\,T^+(\th_1)_a^m\;,
\eas
for generic rapidities $\th_1,\th_2$. Finally we require that for
fixed $\th_1$ the operator $C_{ab}(\th_1,\th_2)$ is meromorphic in
$\th_2$ with a simple pole at $\th_2=\th_1+i\pi$; the residue being
given by
\vspace{4mm}
\eql{C5}
\bas \jot5mm
&& 2\pi i\,\mbox{res}\,C_{ab}(\th,\th+i\pi) = C_{ab}\;,
\eas
which is consistent with (C3).

Implicit in this definition, of course, is the presupposition that
by means of (C1) -- (C4), the prescription (E) is consistent with
the relations of the algebra $F_*(S)$. The consistency of (E) with
$F_*(S)$ imposes the following conditions on $E^{\pm}_{ab}(\th_1,\th_2)$
\ba
E^{\pm}_{ab}(\th_1,\th_2)\is S_{ab}^{dc}(\th_{12})\,
E^{\mp}_{cd}(\th_2,\th_1)\;,\;\;\;Re\,\th_{12}\neq 0\;,\\
S_{ab}^{mn}(\th_{02})\,E^+_{cn}(\th_1,\th_2)\,
W_m(\th_0) \is S_{ca}^{mn}(\th_{10})
W_n(\th_0)\,E^+_{mb}(\th_1,\th_2)\;,\\
S_{ab}^{mn}(\th_{02})\,E^+_{cn}(\th_1,\th_2)\,
T^{\pm}(\th_0)_m^d \is S_{ca}^{mn}(\th_{10})
T^{\pm}(\th_0)_n^d\,E^+_{mb}(\th_1,\th_2)\;,\\
sT^-(\th_1)_a^m\, E^+_{mb}(\th_1 +2\pi i,\th_2) \is
E^-_{bm}(\th_2,\th_1)\, sT^+(\th_1)_a^m\;.
\ea
In Eqn.s (B.3), (B4) one can also replace $E^+_{ab}$ by $E^-_{ab}$,
the replacement being consistent with (B.2). The origin of (B.2) -- (B.4)
is obvious. The last condition is needed to ensure the consistency of
(S) and (E) or equivalently of (2.7) and (E). Eqn. (B.5) arises from
rearranging $W_a(\th_1 +2\pi i) W_b(\th_2) =E^+_{ab}(\th_1 +2\pi i,\th_2)$
and  $W_a(\th_1) W_b(\th_2+2\pi i) =E^-_{ab}(\th_1,\th_2+2\pi i)$
for $\pm Im\,\th_{12}>0$, respectively. Notice that one can only shift
the first, respectively second, argument in $W_a(\th_1)W_b(\th_2)$ by
$2\pi i$ without violating the condition $\pm Im\,\th_{12}>0$. As a
consequence, the reversed Eqn.s
$sT^-(\th)_a^m\, E^-_{mb}(\th_1 +2\pi i,\th_2)=
E^+_{bm}(\th_2,\th_1)\, sT^+(\th_1)_a^m$ do not arise as consistency
conditions. It will become clear below that (B2) -- (B5) are in fact
a complete set of consistency conditions. One can now check from the
definitions (B1) that these Eqn.s hold by means of
(C1) -- (C4). Of course, one still has to check that (C1) -- (C4)
are consistent with $F_*(S)$. Explicitely, this means that the
the consistency  conditions arising from "pushing the
$T^{\pm}(\th)_a^b$ or the $W_a(\th)$ generators through the relations
(C1)--(C4)" all are identities within $F(S)$. For the $W$-generators
this is trivial due to (C1). For the $T(S)$ generators it is easy
to see that it suffices to push $T^{\pm}(\th)_a^b$ through the
relations (C3) and (C4). Let us illustrate the procedure for (C4).
For the l.h.s one finds
\bas
&& T^{\pm}(\th_3)_c^d\,T^+(\th_1)_a^m C_{mb}(\th_1,\th_2)
=S_{ca}^{kl}(\th_{31})S_{eb}^{dq}(\th_{32}+2\pi i)\,
T^+(\th_1)_l^p C_{pq}(\th_1,\th_2)T^{\pm}(\th_3)_k^e\;,
\eas
using (T1) and (C3), while the r.h.s. gives
\bas
&& T^{\pm}(\th_3)_c^d\,C_{bm}(\th_2,\th_1+2\pi i)T^-(\th_1)_a^m\nonum
&&\sspace =S_{ca}^{kl}(\th_{31})S_{eb}^{dq}(\th_{32}+2\pi i)\,
C_{qp}(\th_2,\th_1+2\pi i)T^-(\th_1)_l^p\,T^{\pm}(\th_3)_k^e\;.
\eas
Using now (C4) again on the r.h.s one obtains an identity. The consistency
of (C3) is verified similarly. The compatibility of (C1) -- (C4) with
$F_*(S)$ also guarantees that no new identities arise from (B2) -- (B5)
as consistency conditions. Finally consider the residue equation (C5).
It implies
\be
2\pi i\,\mbox{res}\,[W_a(\th+i\pi)W_b(\th)]=
2\pi i\,\mbox{res}\,E^+_{ab}(\th+i\pi,\th)
= L^+_{ab}(\th)-C_{ab}\;,
\ee
where $L^+_{ab}(\th)$ is defined as in (2.11). The second residue Eqn.
in (2.15) is only indirectly available. In order to be consistent with
(C3), the residue $\mbox{res} C_{ab}(\th +i\pi,\th)$ has to be a
non-central operator. The resulting expression for
$2\pi i\,\mbox{res}E^-_{ab}(\th,\th+i\pi)$ has the same transformation
properties as $L^-_{ab}(\th)-C_{ab}$, but does not coincide with it.
In principle one also has to check the consistency of the relations
arising from (B2) -- (B4) by taking the residue at $\th_{12} =\pm i\pi$.
The conditions arising are however just those that guarantee the
consistency of the algebra $R(S)$ and have been checked before.
In summary,
this shows that $F(S)$ is a consistently defined associative extension of
$F_*(S)$. By means of (B.6) and (3.21) it implements the kinematical
residue axiom, and hence may serve as an alternative form factor algebra.

As an aside we remark that the equations (B.2) -- (B.5) have an
interesting interplay with the anti-homomorphism $\sbar$: One checks from
(2.16) and (B.3), (B.4) that
\be
\sbar W_a(\th_0)\, E^{\pm}_{bc}(\th_1,\th_2)
 =  E^{\pm}_{bc}(\th_1,\th_2) \sbar W_a(\th_0)\;,
\ee
consistently for both the expression of the r.h.s of (2.16b). Suppose then
that $\sbar$ has a consistent action on $E^{\pm}_{ab}(\th_1,\th_2)$ and
define
\be
\tau(E^+_{ab})(\th_1,\th_2) := T^-(\th_1)_a^m\,
\sbar E^+_{nm}(\th_2,\th_1 -2\pi i)\,T^+(\th_2)_b^n\;.
\ee
Then $\tau(E^+_{ab})(\th_1,\th_2)$ again satisfies (B.2) -- (B.5).
\pagebreak

\newappsection{Reduction of \mbox{\boldmath $F(S)$} to
\mbox{\boldmath $R(S)$}}

\noindent The algebra $F(S)$ can be related to $R(S)$ and hence to
the ZF-algebra by a simple reduction prescription. To formulate this
reduction note that for $Re\,\th_{12}\neq 0$ the (WW) relations and
(B.2) imply
\be
W_a(\th_1)\,W_b(\th_2) = S^{dc}_{ab}(\th_{12})\;
W_c(\th_2)\,W_d(\th_1) \pm 2E^{\pm}_{ab}(\th_1,\th_2)\;,
\ee
for $\pm Re\,\th_{12} >0$. Equivalently, one can define the normal
product
$$
N[W_a(\th_1)W_b(\th_2)]=
W_a(\th_1)W_b(\th_2)- E^{\pm}_{ab}(\th_1,\th_2)\;,
\;\;\;\pm Re\,\th_{12} \geq 0\;,
$$
and rewrite (B.9) as
\be
N[W_a(\th_1)\,W_b(\th_2)] = S^{dc}_{ab}(\th_{12})\;
N[W_c(\th_2)\,W_d(\th_1)]\;.
\ee
By definition $N[W_a(\th_1)W_b(\th_2)]$ can be taken to be regular for
fixed $\th_1$ and $0\leq Im\,\th_2 \leq \pi$.
Consider now the reduction of $F(S)$ induced by the following
prescription
\be
C_{ab}(\th_1,\th_2) \rra \pi C_{ab}\,\delta(\th_{12}\pm i\pi)\;,
\ee
where $\delta(\th)$ is again $\delta(Re\,\th)\,\delta_{Im\,\th,0}$.
One can easily verify that the r.h.s. is a solution of
(C1)--(C3). The pole singularities in (C5) get replaced by the
delta distribution singularities at the same positions. Doing this,
the extra conditions (C4) can be dropped, provided the relations (S) are
changed into (A.30). The relations (B.10) then translate into (A.29), the
(TW) relations remain unaffected. Taken together, this means that
the form factor algebra $F(S)$ reduced by the prescription (B.11)
is isomorphic to the
algebra $F_R(S)$ introduced in section A.5. For rapidities that are
real modulo $i\pi$, the latter in turn has been seen to be isomorphic to
the real rapidity algebra $R(S)$. Finally, the reduction
$T^+(\th)_a^b=-T^-(\th)_a^b$ leads to an algebra
$TZ(S)$ which is a semidirect product of $Z(S)$ and $T(S)$. In summary
the relations among the various algebras are
\bas
&& F(S)\;\stackrel{red.}{\rra}\;F_R(S)\,\simeq \,R(S)\;
 \stackrel{red.}{\rra}\;TZ(S)\;\stackrel{subalg.}{\supset}Z(S)\;.
\eas

\pagebreak

\noindent{\tt Acknowledgements:}
I wish to thank L. Faddeev, P. Weisz and J. Balog for instructive
discussions; J. Balog in particular for emphazising the need
to understand the relation between the real and the complex
rapidity situations. Also I thank H.~Ewen and O.~Ogievetsky
for some useful comments.
\vspace{1cm}


\end{document}